\documentclass[journal]{IEEEtran}
\usepackage{bm}
\usepackage{graphicx}
\usepackage{multirow}
\usepackage{amsopn, amsthm,amsmath,amssymb}
\usepackage{mathrsfs}
\usepackage{cite}

\usepackage{latexsym}
\usepackage{rotating}

\usepackage{color}
\usepackage{booktabs}
\usepackage{lineno}

\newcommand{\myref}[1]{Eq. \ref{#1}}

\begin{document}

\title{Weakly-Supervised Video Moment Retrieval via \\ Regularized Two-Branch Proposal Networks with \\ Erasing Mechanism}

\author{
	Haoyuan~Li,
        Zhou~Zhao,
        Zhu~Zhang,
        Zhijie~Lin
        
\thanks{H. Li, Z. Zhao, Z. Zhang and Z. Lin are with the College of Computer Science, Zhejiang University, Hangzhou 310027, China (e-mail: lihaoyuan@zju.edu.cn, zhaozhou@zju.edu.cn; zhangzhu@zju.edu.cn; linzhijie@zju.edu.cn).}
}



\markboth{IEEE Transactions on Image Processing 2021}%
{IEEE Transactions on Image Processing 2021}

\maketitle

\begin{abstract}
Video moment retrieval is to identify the target moment according to the given sentence in an untrimmed video. Due to temporal boundary annotations of the video are extremely time-consuming to acquire, modeling in the weakly-supervised setting is increasingly focused, where we only have access to the video-sentence pairs during training.
Most existing weakly-supervised methods adopt a MIL-based framework to develop inter-sample confrontment, but neglect the intra-sample confrontment between moments with similar semantics. Therefore, these methods fail to distinguish the correct moment from plausible negative moments. 
Further, the previous attention models in cross-modal interaction tend to focus on a few dominant words exorbitantly, ignoring the comprehensive video-sentence correspondence.
In this paper, we propose a novel Regularized Two-Branch Proposal Network with Erasing Mechanism to consider the inter-sample and intra-sample confrontments simultaneously. Concretely, we first devise a language-aware visual filter to generate both enhanced and suppressed video streams. Then, we design the sharable two-branch proposal module to generate positive and plausible negative proposals from the enhanced and suppressed branch respectively, contributing to sufficient confrontment. 
Besides, we introduce an attention-guided dynamic erasing mechanism in enhanced branch to discover the complementary video-sentence relation.
Moreover, we apply two types of proposal regularization to stabilize the training process and improve model performance. The extensive experiments on ActivityCaption, Charades-STA and DiDeMo datasets show the effectiveness of our method.
\end{abstract}
%
%

%
\IEEEpeerreviewmaketitle

\section{Introduction} 
\IEEEPARstart{V}{ideo} 
moment retrieval is a novel topic in information retrieval systems, which integrates computer vision and natural language processing. Given an untrimmed video and a natural language description, video moment retrieval~\cite{gao2017tall,hendricks2017localizing} aims to localize temporal boundaries of the target moment semantically matching to the sentence. 
As shown in Fig. 1, the sentence describes complicated events and corresponds to a temporal moment with complex object interactions. 
Recently, a large number of approaches~\cite{gao2017tall,hendricks2017localizing,chen2018temporally,zhang2019cross-modal,wang2019language-driven} have been proposed to handle this challenging task and have achieved satisfactory performance. 
However, most existing methods are trained in fully-supervised setting. 
Such manual annotations are extremely expensive and time-consuming, especially for ambiguous descriptions. But actually there are a mass of coarse video descriptions on the Internet, such as the captions for videos on YouTube. Hence, we propose a weakly-supervised method for video moment retrieval in this paper, which only needs the video-level sentence annotations rather than temporal alignment annotations for each sentence during training. 

\begin{figure}[t]
	\centering
	\includegraphics[width=1.0\columnwidth]{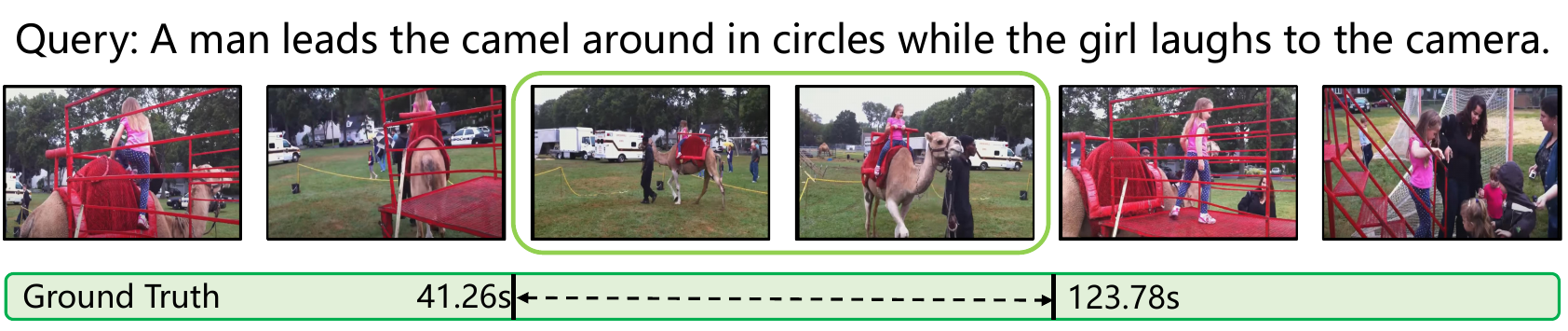}
	\caption{An example of video moment retrieval.}
	\label{fig:example1}
\end{figure}

Most existing weakly-supervised moment retrieval works~\cite{mithun2019weakly,gao2019wslln,chen2020look} adopt a Multiple Instance Learning (MIL)~\cite{karpathy2015deep} based methods. They regard matched and unmatched video-sentence pairs as positive and negative samples respectively. Next, they capture the latent visual-textual alignment by inter-sample confrontment and utilize intermediate results to retrieve the target moment. 
Concretely, Mithun et al.~\cite{mithun2019weakly} apply the text-guided attention weight across video frames to decide the target moment. 
And Gao and Chen et al.~\cite{gao2019wslln,chen2020look} measure the semantic consistency between videos and sentences and utilize video segment scores as localization clues directly. 
However, these methods mainly concentrate on the inter-sample confrontment to judge whether the video matches with the given language descriptions, neglecting the intra-sample confrontment to determine which moment matches the given sentence best. 
Specifically, as Fig. 1 shown, given a matched video-sentence pair, the video generally contains consecutive and complicated contents and these are a number of plausible negative moments, which have a bit of relevance to the sentence. It is intractable to distinguish the correct moment from plausible negative moments, especially when these plausible ones overlap with the ground truth in quantity. Therefore, we need to develop intra-sample confrontment sufficiently between moments with similar information in a video.

Further, for existing MIL-based methods, the cross-modal attention distribution tends to be sharp and a few dominant words are focused exorbitantly as shown in Fig. 2, only the moment corresponding to these words can be localized. 
Though these methods already achieve the inter-sample confrontment, they fail to capture the complete language semantics and comprehensive visual-textual correspondences for moment retrieval. Thus, these previous methods tend to localize the incomplete moment by partial query semantics. 
Actually, the similar problem\cite{Wang2017A-Fast-RCNN,Singh17Hide,Wei17Object,ChoeS2019Attention-Based,Zhang18Adversarial} has arisen before in visual domains(e.g. object localization, semantic segmentation), networks are only responsive to small and sparse discriminative regions from the object of interest, leading to incomplete and inaccurate localization. To mine integral regions in an image, researchers apply an erasing method\cite{Wang2017A-Fast-RCNN,Singh17Hide,Wei17Object,ChoeS2019Attention-Based,Zhang18Adversarial} to mask the dominant regions of an image according to the attention heat map. It forces the network to notice the rest area of the image, improving the model robustness. As for video moment retrieval, we also introduce erasing mechanism to capture the comprehensive query information and complementary cross-modal relations and boost the performance. By erasing the most focused words, the model is required to concentrate on the other discriminative part of the sentence to develop sample confrontments. Then we refine the proposal score distribution based on erasing network and retrieve the moment by comprehensive query semantics. Further than the most of previous erasing methods, we perform erasing during both training and inference to enhance the robustness and design a reconstruction task, introducing a similarity loss function for better retrieval from erasing network.

\begin{figure}[t]
	\centering
	\includegraphics[width=1.0\columnwidth]{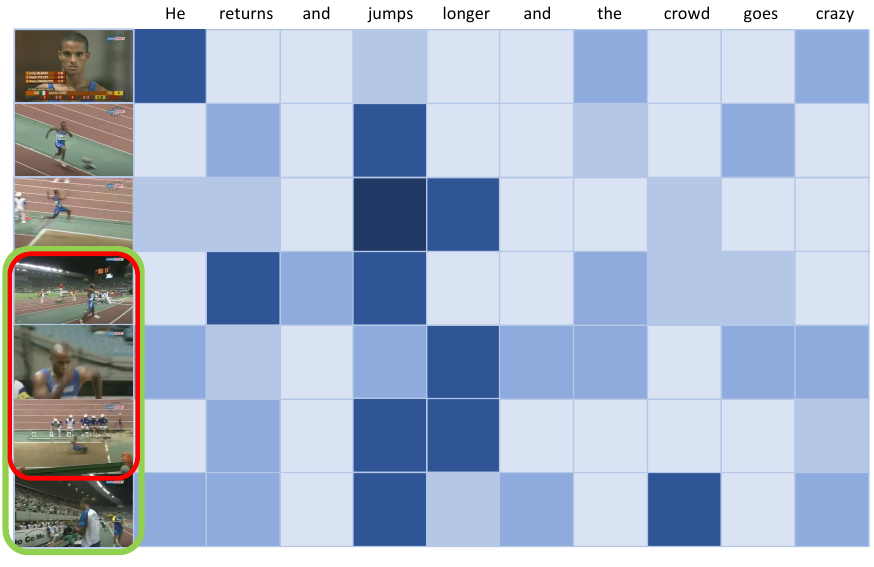} 
	\caption{Cross-Modal Attention Heat Map (The heavier color means the higher correlation of the corresponding frame and word. And the green and the red box on frames represent the ground-truth and the result from our model without erasing respectively).}
	\label{fig:attention}
\end{figure}

In this paper, we extend our previous work
\footnote[1]{This work is the extension of our previous paper\cite{zhang20regularized}, which is accepted by MM(20). The added benefits of the journal paper are clearly and concisely explained in a cover letter that accompanies the submission. And the previous paper is submitted as a supporting document.}
and propose a novel Regularized Two-Branch Proposal Network with Erasing Mechanism (RTPEN) to develop the fine-grained intra-sample confrontment by discovering the plausible negative moment proposals and capture more complete visual-textual relations by attention-guided dynamic erasing. 
Concretely, we first devise a language-aware visual filter to generate an enhanced video stream and a suppressed stream from the original video stream. In the enhanced stream, we highlight the critical frames according to the language information and weaken unnecessary ones. On the contrary, the crucial frames are suppressed in the suppressed stream. 
Next, we design a primary two-branch proposal module with to generate moment proposals from each stream, where the enhanced and suppressed branch produces primary positive and plausible negative moment proposal respectively. 
Besides, in the enhanced branch, we apply an attention-guided dynamic erasing module to erase the dominant words and gain an erased query, then we feed this new positive video-sentence pair into our model again and acquire the complementary moment proposal scores to refine the primary positive score distribution by weighted summation. Thus, we can obtain positive moment proposal based on comprehensive query semantics and cross-modal correspondences. 
During erasing, we also design a reconstruction task, introducing a loss function which is the normalized cosine similarity between erased word features and cross-modal fusion features to inject erased words semantics into the erasing enhanced branch as well as close the cross-modal semantic distance.
By the sufficient confrontment between two branches and the complementary video-sentence relation captured, we can retrieve the most relevant moment from plausible ones accurately. 
However, the suppressed branch may generate simple negative proposals rather than plausible ones, resulting in ineffective confrontment. To avoid it, we share all parameters between two branches to make them possess the same ability to produce high-quality proposals. Moreover, parameter sharing can reduce network parameters and accelerate model convergence. By the sharable two-branch framework, we can develop sufficient inter-sample and intra-sample confrontments simultaneously to boost the performance of weakly-supervised video moment retrieval.

Next, we consider the concrete design of the language-aware visual filter and two-branch proposal module with erasing mechanism. For the language-aware visual filter, we first project the textual features into fixed cluster centers by a trainable generalized Vector of Locally Aggregated Descriptors (VLAD)~\cite{arandjelovic2016netvlad}, where each cluster center can be regarded as a language scene, and then we calculate attention scores between the scene and frame features as language-to-frame relevance. Such a scene-based method can introduce an intermediately semantic space for sentences and videos, contributing to boosting the generalization ability. Next, we apply a min-max normalization on the distribution to avoid producing a trivial score distribution, e.g. all frames are assigned to 1 or 0. Based on the normalized distribution, we employ a two-branch gate to generate the enhanced and suppressed streams. 

As for the two-branch proposal module, two branches have the consistent structure and share parameters. We first apply a conventional cross-modal interaction method~\cite{chen2018temporally,zhang2019cross} between frame sequences and language with dynamic erasing mechanism.
Next, we apply a two-dimensional temporal map~\cite{zhang2019learning} to capture relations between adjacent video moments. After it, we need to generate high-quality moment proposals from the enhanced branch. Most existing weakly-supervised methods~\cite{mithun2019weakly,gao2019wslln,chen2020look} take all frames or moments as proposals to develop the inter-sample confrontment, which introduces a large amount of ineffective proposals during training.
Different from them, we suggest a center-based proposal method to filter out unnecessary proposals and retain high-quality ones only. Specifically, we first determine the moment with the highest score as the center and then select those moments having high overlaps with the center one. This technique can effectively select a series of correlative moments to make the confrontment between two branches more sufficient.

Network regularization is a widely-used technique in weakly-supervised tasks~\cite{liu2019completeness,chen2019weakly}, which injects extra limitations~(i.e. prior knowledge) into the network to stabilize the training process and improve the model performance. Here we design a proposal regularization strategy for our model, consisting of a global term and a gap term.
On the one hand, considering most of moments are semantically irrelevant to the language descriptions, we apply a global regularization term to make the average moment score relatively low, which implicitly encourages the scores of irrelevant moments close to 0.
On the other hand, we further expect to select the most accurate moment from positive moment proposals, thus we apply another gap regularization term to enlarge the score gaps between those positive moments for better identifying the target one.

Our main contributions are as follows:
\begin{itemize}
	\item We design a novel Regularized Two-Branch Proposal Network with Erasing Mechanism (RTPEN) for weakly-supervised video moment retrieval, considering the inter-sample and intra-sample confrontments simultaneously by the sharable two-branch framework.
	\item We devise the language-aware visual filter to produce the enhanced and suppressed video streams, and develop the sharable two-branch proposal module to generate the positive moment proposals and plausible negative ones for sufficient intra-sample confrontment.
	\item We introduce an attention-guided dynamic earsing mechanism to encourage the model to discover comprehensive latent visual-textual alignments for retrieval. 
	\item We apply two proposal regularization strategies to stabilize the training process and boost the model performance.
	\item The sufficient experiments on three large-scale datasets demonstrate the effectiveness of our proposed RTPEN model.
\end{itemize}

The organization of this paper is as follows. Section II introduce the related work in temporal action detection, video moment retrieval and erasing approach. Next the detailed architecture of RTPEN model is described in Section III. Then, Section IV shows the experiments on ActivityCaption\cite{regneri2013grounding}, Charades-STA\cite{gao2017tall} and DiDeMo\cite{hendricks2017localizing} comparing with both fully-supervised and weakly-supervised models. Finally, the conclusion is illustrated in Section V.


\section{Related Work}
In this section, we briefly review some related works on temporal action detection, video moment retrieval and erasing approach.

\subsection{Temporal Action Detection}
Temporal action detection aims to localize the temporal boundaries as well as the categories of action instances in untrimmed videos. From the perspective of our interest, the works pertaining to temporal action detection can be categorized as either fully-supervised or weakly-supervised. Most fully-supervised methods~\cite{shou2016temporal,zhao2017temporal,shou2017cdc,chao2018rethinking,zeng2019graph, Xu2020Sub} apply the two-stage framework, which first generates a series of temporal action proposals, then predicts the action category and regresses their boundaries. Concretely, Shou et al.~\cite{shou2016temporal} design three segment-based 3D ConvNets, which are proposal, classification and localization networks, to accurately detect action instances and Zhao et al.~\cite{zhao2017temporal} apply a structured temporal pyramid to model the temporal structure of each action. Recently, Chao et al.~\cite{chao2018rethinking} transfer the classical Faster-RCNN object detection framework~\cite{ren2015faster} for action detection, Zeng et al.~\cite{zeng2019graph} and Xu et al.~\cite{Xu2020Sub} use graph convolutional networks to exploit proposal-proposal relations and find the effective video context. 

Under the weakly-supervised setting only with coarse video-level action labels, Wang et al.~\cite{wang2017untrimmednets} introduce the classification and selection modules to learn the action models and reason about the temporal duration of action instances. Nguyen et al.~\cite{nguyen2018weakly} design the model to identify a sparse subset of key segments associated with target actions in a video utilizing an attention module and fuse the key segments through adaptive temporal pooling. Further, Shou et al.~\cite{shou2018autoloc} propose a novel Outer-Inner-Contrastive loss to automatically discover the segment-level supervision for action boundary prediction. To keep the completeness of actions, Liu et al.~\cite{liu2019completeness} employ a multi-branch network where branches are enforced to discover distinctive action parts. And Yu et al.~\cite{yu2019temporal} explore the latent temporal action structure and model each action instance as a multi-phase process.

\begin{figure*}[t]
	\centering
	\includegraphics[width=2.0\columnwidth]{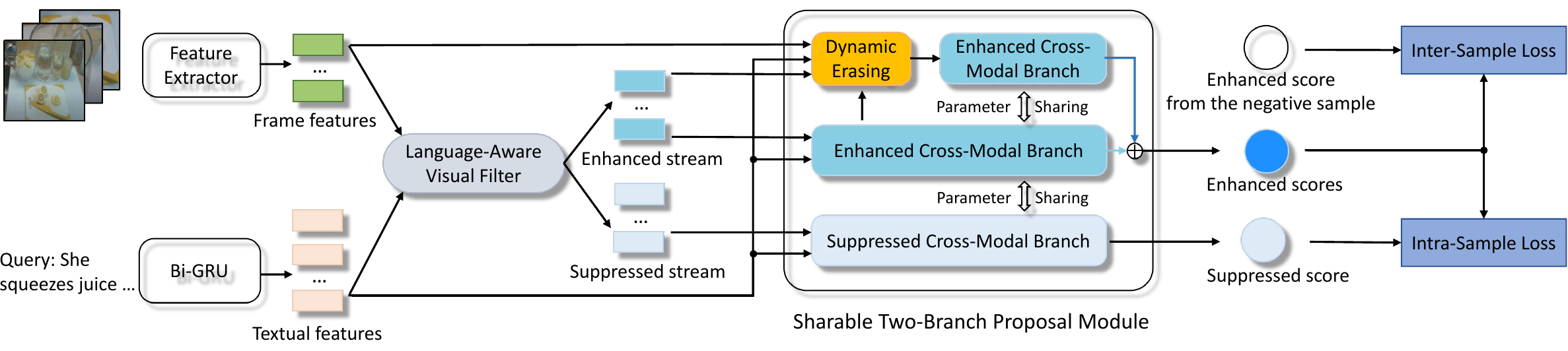} 
	\caption{The Overall Architecture of the Regularized Two-Branch Proposal Network with Erasing Mechanism (RTPEN). 
	}
	\label{fig:framework}
\end{figure*}

\subsection{Video Moment Retrieval}
Video moment retrieval aims to identify the target moment in untrimmed videos according to a given sentence query. Most existing approaches employ the top-down framework, which first sets a series of moment proposals and then selects the most relevant, tackling moment retrieval as a ranking problem.
Early methods~\cite{gao2017tall,hendricks2017localizing,hendricks2018localizing,liu2018attentive,liu2018cross}  generate moment proposals by sliding windows with various lengths explicitly and calculate the correlation of the query and each proposal in a cross-modal space individually. To incorporate long-term video context and captures the long-range semantic dependency, researchers~\cite{chen2018temporally,zhang2019man,zhang2019cross,lin2020moment,xu2019multilevel,zhang2019learning,yuan2019semantic} produce moment proposals implicitly by defining multiple temporal anchors after visual-textual interactions.
Concretely, Chen et al.~\cite{chen2018temporally} build fine-grained frame-by-word interactions and aggregate the matching clues dynamically.
Zhang et al.~\cite{zhang2019man} build a structured graph to model moment-wise temporal relations and propose an iterative graph adjustment network to learn the best structure. 
And Yuan et al.~\cite{yuan2019semantic} introduce a semantic conditioned dynamic modulation model with respect to diverse video contents, correlating and composing the sentence-related video information over time. 
Further, Zhang et al.~\cite{zhang2019learning} design a two-dimensional temporal map to capture the temporal relations between adjacent video moments.
Different from the top-down formula mentioned above, the bottom-up framework~\cite{chen2019localizing,chenrethinking} is to predict the probability of each frame as the target boundary directly. 
He and Wang et al.~\cite{he2019read,wang2019language-driven} formulate moment retrieval as a sequential decision making problem and apply the reinforcement learning to regulate temporal boundaries progressively. 
Besides temporal video moment retrieval, recent works~\cite{chen2019weakly,zhang2020does,zhang2020object} ground the spatio-temporal tubes as well from videos according to the give sentences. Addictionally, Zhang et al.~\cite{zhang2019localizing} localizealize the temporal moment by the image query rather than the natural language query.

Recently, researchers~\cite{duan2018weakly,mithun2019weakly,gao2019wslln,lin2019weakly,chen2020look} start to explore the moment retrieval in weakly-supervised setting, only with video-level annotations. 
Duan et al.~\cite{duan2018weakly} regard video event captioning and moment retrieval as dual problems, and develop a cycle system to train models. 
Mithun, Gao and Chen et al.~\cite{mithun2019weakly,gao2019wslln,chen2020look} apply a MIL-based framework to capture latent visual-textual relation by inter-sample confrontment. 
Mithun et al.~\cite{mithun2019weakly} introduce a joint multi-modal embedding based framework and determine the reliant moment utilizing Text-Guided Attention. 
Gao et al.~\cite{gao2019wslln} propose an alignment module to measure the semantic consistency between videos and sentences, and devise a detection module to compare the moment proposals. 
And Chen et al.~\cite{chen2020look} utilize a two-stage model to detect the moment in a coarse-to-fine manner. 
Besides MIL-based methods, Lin et al.~\cite{lin2019weakly} design a semantic completion network to rank moment proposals by a sentence reconstruction reward, but ignore intra-sample confrontment.
Unlike previous methods, we design a sharable two-branch framework with erasing mechanism to consider the inter-sample and intra-sample confrontments simultaneously for weakly-supervised moment retrieval.

\subsection{Erasing Approach}
Erasing is an effective data augmentation method to suppress over-fitting and enhance the robustness of models. Previous works has explored erasing a lot in visual domain, such as object detection\cite{Wang2017A-Fast-RCNN,Singh17Hide}, semantic segmentation\cite{Wei17Object} and action localization\cite{Singh17Hide}. There are several erasing strategies, the most classic is random erasing\cite{Singh17Hide, ChoeS2019Attention-Based}, they randomly select regions in an image and replace their pixels with 0 or mean value of the image, producing a multitude of new images for training. Besides, Zhang et al. propose a multi-branch\cite{Zhang18Adversarial} erasing architecture, where each branch is forced to mask different regions then fuse every branch to generate complementary image heat map. Further, Wei et al. utilise step-wise\cite{Wei17Object, En19Human} strategy to erasing image, the common approach is taking gradients or attention scores by the whole image as a guidance to erase the dominant regions and encourage the model to focus on the comprehensive regions.

Close to our task, in weakly-supervised video action detection, Singh et al.\cite{Singh17Hide} randomly erase frames segments during training to force the network to learn the relevant frames corresponding to an action. In cross-modal setting, Liu et al.\cite{Liu19Improving} take a step further towards erasing in both images and sentences in expression grounding task. To the best of our knowledge, we first offer the attempt to extend the adversarial erasing mechanism for video moment retrieval, which helps discover video-sentence alignment and promote the performance of the retrieval.
%
%
%

\begin{figure*}[t]
	\centering
	\includegraphics[width=2.0\columnwidth]{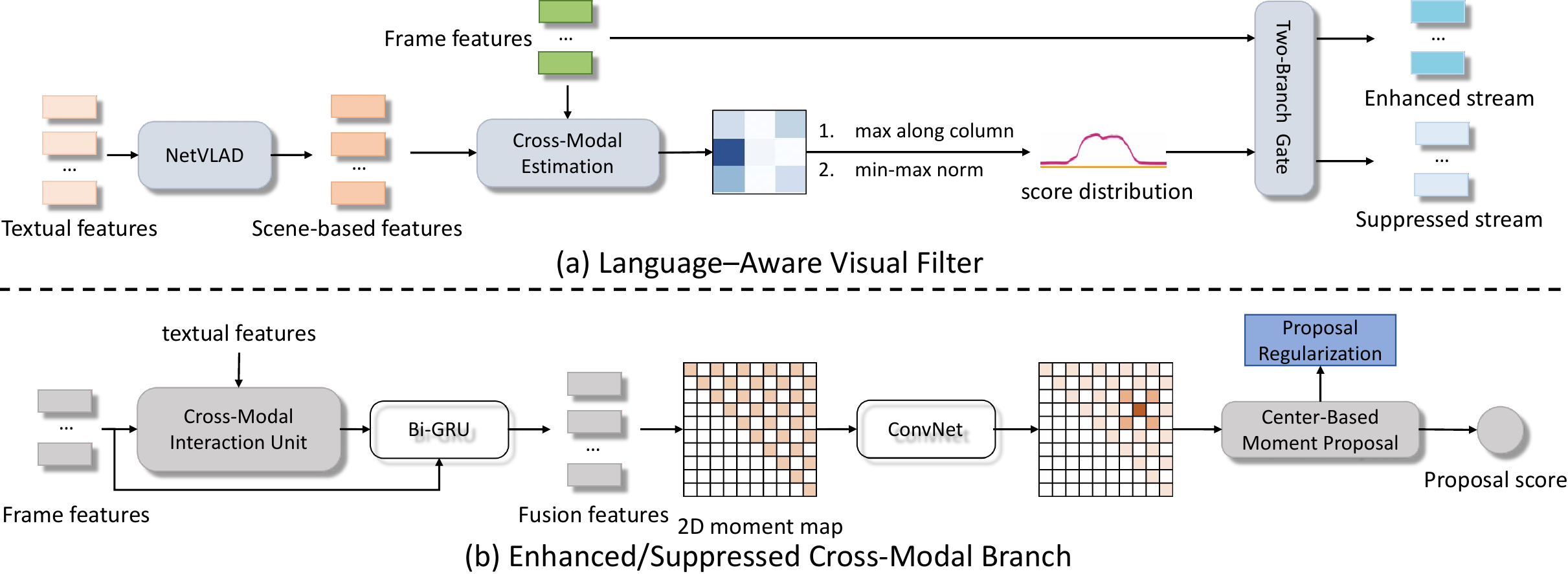} 
	\caption{ The Concrete Designs of the Language-Aware Visual Filter and Primary Sharable Two-Branch Proposal Module. 
	}
	\label{fig:framework2}
\end{figure*}

\section{The Proposed Method}
Given a sentence query $Q$, video moment retrieval aims to detect the most relevant moment ${\hat l} = ({\hat s},{\hat e})$ within a video $V$, where ${\hat s}$ and ${\hat e}$ denote the coordinates of the start and end frames of the target moment. Due to the weakly-supervised setting, we only utilize the coarse video-level annotations.

\subsection{The Overall Architecture Design}
We first introduce the overall architecture design of our Regularized Two-Branch Proposal Network with Erasing Mechanism (RTPEN). 
As Fig.~\ref{fig:framework} shown, we devise a language-aware visual filter to produce the enhanced and suppressed video streams, and next propose the sharable two-branch proposal module with dynamic erasing mechanism to generate the positive and plausible negative moment proposals. Finally, we develop the inter-sample and intra-sample losses with two types of proposal regularization terms.

Concretely, we first extract word features of the sentence query $Q$ by a pre-trained Glove embedding~\cite{pennington2014glove}. We then feed the word features into a Bi-GRU ~\cite{chung2014empirical} to learn word semantic representations ${\bf Q} = \{{\bf q}_i\}_{i=1}^{n_q}$, where $n_q$ is the word number and ${\bf q}_i$ is the semantic feature of the $i$-th word aggregating the contextual information. As for the video $V$, we first extract video features using a pre-trained feature extractor (e.g. 3D-ConvNet~\cite{tran2015learning}) and then apply a temporal average pooling to shorten the video sequence. We denote frame features as ${\bf V} = \{{\bf v}_i\}_{i=1}^{n_v}$, where $n_v$ is the number of frames.

After extracting features, we devise a language-aware visual filter to produce the enhanced and suppressed video streams, given by 
\begin{equation}
	{\bf V}^{en}, {\bf V}^{sp} = {\rm Filter} \ ({\bf V}, {\bf Q}),\label{filter}
\end{equation}
where ${\bf V}^{en} = \{{\bf v}^{en}_i\}_{i=1}^{n_v}$ and ${\bf V}^{sp} = \{{\bf v}^{sp}_i\}_{i=1}^{n_v}$ represent the enhanced and suppressed video streams respectively. In the enhance stream, we highlight the significative frame features relevant to the sentence and weaken unnecessary frames. On the contrary, the critical frames are suppressed in the suppressed stream.

Next, we propose a sharable two-branch proposal module with dynamic erasing mechanism to produce the positive and plausible negative moment proposals. This module consists of an enhanced branch and a suppressed branch, where the enhanced branch is called twice for primary and erasing cross-modal interaction. For more robust and lighter model, two branch have the consistent structure and share parameters ${\Theta}$, given by
\begin{eqnarray}
	\begin{aligned}
		& {\rm P}^{en}, {\bf C}^{en} =  {\rm EnhancedBranch}_{\Theta} \ ({\bf V}^{en}, {\bf Q}), \\
		& {\rm P}^{sp}, {\bf C}^{sp} =  {\rm SuppressedBranch}_{\Theta} \ ({\bf V}^{sp}, {\bf Q}), \\ 
	\end{aligned}
	\label{2branch}
\end{eqnarray}
we feed the enhanced video stream ${\bf V}^{en}$ and textual features ${\bf Q}$ into the enhanced branch and generate the positive moment proposals ${\rm P}^{en} = \{{p}^{en}_i\}_{i=1}^{T}$ and their scores ${\bf C}^{en} = \{{c}^{en}_i\}_{i=1}^{T}$, where $T$ is the number of moment proposals, and each proposal ${p}^{en}_i$ corresponds the start and end timestamps $({s}^{en}_i, {e}^{en}_i)$ and a confidence score ${c}^{en}_i \in (0, 1)$. Likewise, the suppressed branch produces ${\rm P}^{sp}$ and ${\bf C}^{sp}$ from the suppressed video stream. 
Further, we introduce the dynamic erasing mechanism in the enhanced proposal branch, utilizing the attention distribution computed in primary enhanced branch as the guidance to mask top words ${{\bf W}^*} = \{{{\bf w}_i^*}\}_{i=1}^{n_{e}}$ and gain a new sentence ${\bf Q}^* = \{{\bf q}_i^*\}_{i=1}^{n_q}$, where $n_e$ is the number of erased words. After that, we can generate a new enhanced stream ${\bf V}^{en^*} = \{{\bf v}^{en^*}_i\}_{i=1}^{n_v}$ and proposal scores ${\bf C}^{en^*} = \{{c}^{en^*}_i\}_{i=1}^{T}$ following the \myref{filter} and \myref{2branch} respectively. 
By ${\bf C}^{en^*}$, we refine ${\bf C}^{en}$ to ${\bf C}^{en^f}$, and calculate the enhanced score $K^{en} = \sum_{i=1}^{T}{c}^{en^f}_i$ and suppressed score $K^{sp} = \sum_{i=1}^{T}{c}^{sp}_i$. The intra-sample loss is given by
\begin{equation}
	{\mathcal L}_{intra}  = {\rm max}(0, \ \Delta_{intra} - {K}^{en} + {K}^{sp}),
\end{equation}
where ${\mathcal L}_{intra}$ is a margin-based triplet loss and $\Delta$ is the margin. Due to the parameter sharing between two branches, the suppressed branch will select plausible negative proposals. By the sufficient intra-sample confrontment, we can distinguish the target moment from intractable negative moments.

In addition to the intra-sample loss, we also develop a inter-sample loss by unmatched video-sentence sample, i.e. the negative sample. Specifically, for each sentence query $Q$, we randomly select an unmatched video ${\overline V}$ from the training set to form a negative sample $({\overline V},Q)$. Likewise, we can randomly choose a unmatched query ${\overline Q}$ to construct another negative sample $(V,{\overline Q})$. Next, we apply the RTPEN to obtain the enhanced scores ${\overline K}^{en}_{V}$ and ${\overline K}^{en}_{Q}$ for negative samples. Thus, the inter-sample loss is given by
\begin{eqnarray}
	\begin{aligned}	
	{\mathcal L}_{inter} = \ 
	&{\rm max}(0, \ \Delta_{inter} - {K}^{en} + {\overline K}^{en}_{V})\  + \\
	&{\rm max}(0, \ \Delta_{inter} - {K}^{en} +  {\overline K}^{en}_{Q}),
	\end{aligned}
\end{eqnarray}
where the $\Delta_{inter}$ is the margin and ${\mathcal L}_{inter}$ encourages the enhanced scores of positive samples to be higher than negative ones. 

\subsection{Language-Aware Visual Filter}
We next introduce the language-aware visual filter with the scene-based cross-modal estimation. 
To compute the language-relevant score distribution over frames, we first apply a NetVLAD~\cite{arandjelovic2016netvlad} to project the sentence features ${\bf Q} = \{{\bf q}_i\}_{i=1}^{n_q}$ into several cluster centers. Concretely, given the trainable center vectors ${\bf C} = \{{\bf c}_j\}_{j=1}^{n_c}$ where $n_c$ is the number of cluster centers, the NetVLAD can accumulate the residuals between original language features and center vectors by a soft assignment, given by
\begin{eqnarray}
	& {\bf \alpha}_{i} = {\rm softmax}({\bf W}^c{\bf q}_i+ {\bf b}^c), \
	{\bf u}_{j} =  \sum_{i=1}^{n_q} {\bf \alpha}_{ij}({\bf q}_i - {\bf c}_j), \label{filter1}
\end{eqnarray}
where ${\bf W}^{c}$ and ${\bf b}^{c}$ are projection matrix and bias, and ${\bf \alpha}_{i} \in \mathbb{R}^{n_c}$ produced by softmax operation is the soft assignment coefficients corresponding to $n_c$ centers. The ${\bf u}_{j}$ is the accumulated features from ${\bf Q}$ for the $i$-th center. We can regard each cluster center as a language scene and ${\bf u}_{j}$ is a scene-based language feature. Then, we calculate the cross-modal matching scores between video features $\{{\bf v}_i\}_{i=1}^{n_v}$ and language features $\{{\bf u}_j\}_{j=1}^{n_c}$ by 
\begin{equation}
	\beta_{ij} = \sigma ({\bf w}^{\top}_a{\rm tanh}({\bf W}^{a}_{1} {\bf v}_{i}+ {\bf W}^{a}_{2}{\bf u}_{j}+{\bf b}^{a})),\label{filter2}
\end{equation}
where $ {\bf W}_{1}^{a}$, $ {\bf W}_{2}^{a}$ are projection matrices, ${\bf b}^{a}$ is the bias, ${\bf w}^{\top}_a$ is the row vector and $\sigma$ is the sigmoid function. The $\beta_{ij} \in (0,1)$represents the matching score of the $i$-th frame feature and $j$-th scene-based language feature. That is, the scene-based method introduces an intermediately semantic space for videos and texts.

Considering a frame should be significant if it is associated with any language scene, we compute a holistic score for the $i$-th frame by ${\overline \beta}_{i} = {\rm max}_j\{\beta_{ij}\}$. To avoid producing a trivial score distribution, e.g. all frames are assigned to 1 or 0, we apply a min-max normalization on the distribution by 
\begin{equation}
	{\widetilde \beta}_{i} = \frac{{\overline \beta}_{i} -  \min_{j}\{{\overline \beta}_{j}\}}{\max_j\{{\overline \beta}_{j}\} - \min_j\{{\overline \beta}_{j}\}}.\label{filter3}
\end{equation}
Therefore, we acquire the normalized distribution $\{{\widetilde \beta}_{i}\}_{i=1}^{n_v}$ over frames, which means the relevance between the $i$-th frame and language queries. 
Next, we apply a two-branch gate to produce the enhanced and suppressed streams, denoted by
\begin{equation}
	\begin{aligned}
		&{\bf v}^{en}_i = {\widetilde \beta}_{i} \cdot {\bf v}_i, \ \ {\bf v}^{sp}_i = (1-{\widetilde \beta}_{i}) \cdot {\bf v}_i,\label{filter4}
	\end{aligned}
\end{equation}
where the enhance stream ${\bf V}^{en} = \{{\bf v}^{en}_i\}_{i=1}^{n_v}$ highlights the significative frames and weaken unnecessary ones according to the normalized score, while the suppressed stream ${\bf V}^{sp} = \{{\bf v}^{sp}_i\}_{i=1}^{n_v}$ is the opposite.

\begin{figure}[t]
	\centering
	\includegraphics[width=1.0\columnwidth]{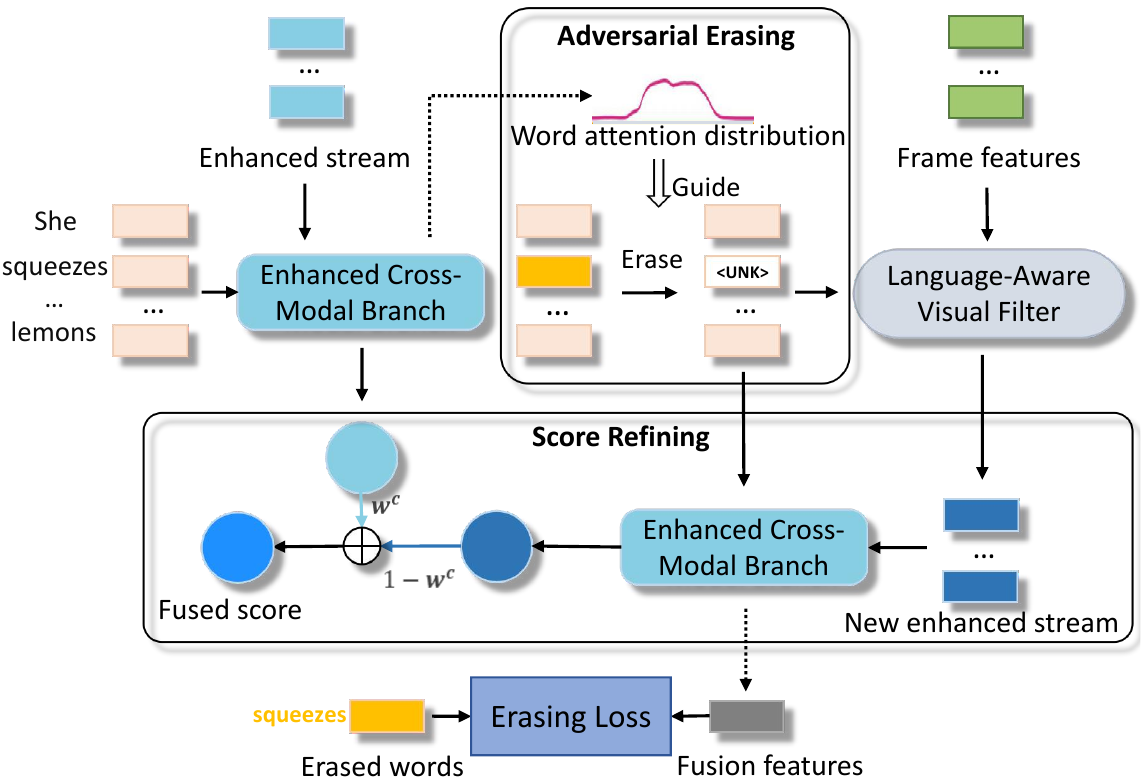} 
	\caption{Dynamic Erasing  Mechanism in Enhanced Branch (Dotted lines represent the intermediate products of modules).}
	\label{fig:framework erase}
\end{figure}

\subsection{Primary Sharable Two-Branch Proposal Module}
In this section, we introduce the primary sharable two-branch proposal module, including an enhanced branch and a suppressed branch with a consistent structure and sharable parameters. The sharing setting can make both branches produce high-quality moment proposals, avoiding the suppressed branch generating too simple negative proposals and leading to the ineffective confrontment.
Here we only present the design of the enhanced branch.

Given the enhanced stream ${\bf V}^{en} = \{{\bf v}^{en}_i\}_{i=1}^{n_v}$ and sentence features ${\bf Q} = \{{\bf q}_i\}_{i=1}^{n_q}$, we first conduct a widely-used cross-modal interaction unit~\cite{zhang2019cross,chen2019localizing} to incorporate textual clues into visual features. Concretely, we perform a frame-to-word attention and aggregate the textual features for each frame, given
\begin{eqnarray}
	\delta_{ij} = {\bf w}^{\top}_m{\rm tanh}({\bf W}^{m}_{1} {\bf v}^{en}_{i}+ {\bf W}^{m}_{2}{\bf q}_{j}+{\bf b}^{m})
	\label{attn}
\end{eqnarray}
\begin{eqnarray}
	{\overline \delta}_{ij} = \frac{{\rm exp}(\delta_{ij} )}{\sum_{k=1}^{n_q} {\rm exp}(\delta_{ik})},  \ {\bf s}^{en}_i = \sum_{j=1}^{n_q} {\overline \delta}_{ij} {\bf q}_{j},
	\label{attn1}
\end{eqnarray}
where ${\bf s}^{en}_{i}$ is the aggregated textual representation relevant to the $i$-th frame.
Then we concatenate two modal features and a Bi-GRU network~\cite{chung2014empirical} is applied to develop the visual-textual interaction sufficiently by
\begin{equation}
	\begin{aligned}
		&{\bf m}_i^{en} = BiGRU([{\bf v}^{en}_i ; {\bf s}_i^{en}]),
		\label{fuse}
	\end{aligned}
\end{equation}
where the ${\bf m}_i^{en}$ is the language-aware frame feature.

Next, we build a 2D moment feature map following the 2D temporal network~\cite{zhang2019learning} and capture relationships between adjacent moments. Specifically, the 2D moment map ${\bf F} \in \mathbb{R}^{n_v \times n_v \times d_m}$ consists of three dimensions: the first two dimensions represent the start and end coordinates of a moment and the third dimension is the fusion feature. We can compute the feature of a moment with temporal duration $[a, b]$  by ${\bf F}[a,b,:] = \sum_{i=a}^{b}{\bf m}^{en}_i$. Note that the location with $a > b$ is invalid and is padded with 0. To avoid too much computational cost, we follow the sparse sampling setting in~\cite{zhang2019learning}. That is, not all moments with $a <= b$ are proposed if the $n_v$ is large. 
With the 2D maps, we conduct the two-layer 2D convolution with the kernel size $K$ to develop moment relationships between adjacent moments. After it, we acquire the cross-modal features $ \{{\bf f}^{en}_i\}_{i=1}^{M_{en}}$, where ${M_{en}}$ is the number of all moments in the 2D map, and calculate their proposal scores $ \{{c}^{en}_i\}_{i=1}^{M_{en}}$  by
\begin{eqnarray}
	{c}^{en}_i = \sigma({\bf W}^p {\bf f}^{en}_i + {\bf b}^p).
	\label{cal prop score}
\end{eqnarray}

Next, we utilize a center-based proposal method to fiter out unnecessary moments and only retain critical ones as the positive moment proposals.
Concretely, we first choose the moment with the highest score ${c}^{en}_i$ as the center moment and rank the rest of moments according to the overlap with the center one. We then select top $T-1$ moments and obtain $T$ positive proposals ${\rm P}^{en} = \{{p}^{en}_i\}_{i=1}^{T}$ with proposal scores ${\bf C}^{en} = \{{c}^{en}_i\}_{i=1}^{T}$. And temporal boundaries $({s}^{en}_i, {e}^{en}_i)$ of each moment are the indices of its location in the 2D map. This method can effectively select a series of correlative moments. Likewise, the suppressed branch has the completely identical structure to generate the plausible negative proposals ${\rm P}^{sp} = \{{p}^{sp}_i\}_{i=1}^{T}$ with proposal scores ${\bf C}^{sp} = \{{c}^{sp}_i\}_{i=1}^{T}$.

\subsection{Dynamic Erasing Mechanism in Enhanced Branch}
After developing primary sharable two-branch proposal module, we introduce an attention guided erasing mechanism in enhanced branch, as shown in Fig.~\ref{fig:framework erase}, to excavate the comprehensive cross-modal correspondence and refine the primary positive proposal scores. First we take out ${\delta}_{ij}$ from \myref{attn}  in primary enhanced proposal module and calculate the textual attention by
\begin{equation}
	{\delta}^*_i = \sum_{j=1}^{n_v}{\delta}_{ij},
\end{equation}
where ${\delta}_i$ is the textual attention, a signal to show how much words are focused. We replace the top percentage $E$ of the most dominant words ${{\bf W}^*}$ in language query by an “Unknown” token and gain a erased sentence ${\bf Q}^{{en^*}}$. 

Then we feed the ${\bf Q}^{en^*}$ and ${\bf V}^{en}$ into the above models by \myref{filter1}--\myref{attn} to obtain a new enhanced stream ${\bf V}^{en^*} = \{{\bf v}^{en^*}_i\}_{i=1}^{n_v}$ and cross-modal attention ${\delta}^*_{ij}$. Similar to \myref{attn1}, we perform a word-to-frame attention and aggregate the visual features for each word given
\begin{equation}
	{\widetilde \delta}^*_{ij} = \frac{{\rm exp}(\delta_{ij} )}{\sum_{k=1}^{n_v} {\rm exp}(\delta^*_{kj})},  \ {\bf s}^{en}_j = \sum_{i=1}^{n_v} {\widetilde \delta}^*_{ij} {\bf v}_{i},
\end{equation}
where ${\bf s}^{en}_{j}$ are the aggregated visual representation relevant to the $j$-th word. 
Specially, we force the model to reconstruct the discarded semantic information by their visual and textual context, introducing an erasing loss given
\begin{equation}
	\begin{aligned}
		&s = \sum_j^{n_q}  cos({{\bf W}^e}{\bf m}_i^{en^*},{\bf q}_j) \quad\ if\  {\bf q}_j\ in\ {{\bf W}^*}, \\
		&{\mathcal L}_{erase} = (s + 1) / 2,
	\end{aligned}
\end{equation}
where ${\bf W}^e$ is the projection matrix, $cos(\cdot)$ is cosine function, which is more stable than dot product with l2 norm\cite{zhang2019learning} during training by our experience. Essentially, the erasing loss is a normalized similarity, which contributes to avoiding losing too much important language information after erasing as well as closing the cross-modal semantic distance. After it, we can obtain the erasing branch proposal scores $ \{{c}^{en^*}_i\}_{i=1}^{M_{en}}$ following \myref{attn1}-\myref{cal prop score}, which helps refine primary branch proposal scores by weighted summation, given
\begin{eqnarray}
	{c}^{en^f}_i = w^c*{c}^{en}_i+(1-w^c)*{c}^{en^*}_i,
\end{eqnarray}
where $w^c$ is a learnable parameter to control which part is more significant, it is initialized to 0.5.

\subsection{Proposal Regularization}
Next, we employ a proposal regularization strategy to inject some prior knowledge into our model, which contains a global term and a gap term.
We only apply the proposal regularization in the enhanced branch due to the parameter sharing between two branches.

Specifically, considering most of moments are unaligned to the language descriptions, we first apply a global term to make the average moment score relatively low, given by 
\begin{equation}
	{\mathcal L}_{global} = \frac{1}{M_{en}}\sum_{i=1}^{M_{en}} {c}^{en^f}_i,
\end{equation}
where $M_{en}$ is the number of all moments in the 2D map.
This global term implicitly encourages the scores of unselected moments in the 2D map close to 0 to stabilize training, while ${\mathcal L}_{intra}$ and ${\mathcal L}_{inter}$ guarantee positive proposals have high scores. 

On the other hand, we further expect to identify the most accurate one as the final localization result from $T$ positive moment proposals, thus it is crucial to enlarge the score gaps between these proposals to make them distinguishable. We perform ${\rm softmax}$ on positive proposal scores and then employ the gap term ${\mathcal L}_{gap}$  by
\begin{equation}
	{\overline c}^{en^f}_i = \frac{{\rm exp}({c}^{en^f}_i)}{\sum_{i=1}^{T}{\rm exp}({c}^{en^f}_i)}, \ \ {\mathcal L}_{gap}  = -\sum_{i=1}^{T} {\overline c}^{en^f}_i  {\rm log}({\overline c}^{en^f}_i), 
\end{equation}
where $T$ is the number of positive proposals rather than the number $M_{en}$ of all proposals.
The ${\mathcal L}_{gap}$ is the comentropy of proposal scores essentially, when it decreases, the score distribution will become more diverse, i.e. it implicitly encourages to enlarge the score gaps between positive moment proposals.

\subsection{Training and Inference}
Based on the aforementioned model design, we apply a multi-task loss to train our RTPEN model in an end-to-end manner, given by
\begin{eqnarray}
	\begin{aligned}
		{\mathcal L}_{RTPEN}\ =\  &{\lambda}_1 {\mathcal L}_{intra} +  {\lambda}_2 {\mathcal L}_{inter}\ + \\ 
		&  {\lambda}_3 {\mathcal L}_{erase} + {\lambda}_4 {\mathcal L}_{global} + {\lambda}_5 {\mathcal L}_{gap} ,
	\end{aligned}
\end{eqnarray}
where ${\lambda}_*$ are the hyper-parameters to control the balance of losses. 

During inference, we can directly select the moment ${p}^{en}_i$  with the highest proposal score ${c}^{en}_i$
which fuses the primary and erasing module scores from the enhanced branch.

\section{Experiments}
In this section, first we describe the datasets explored in experiments and evaluation criteria. Next we introduce the implementation details, then demonstrate the performance of our model and state-of-the-art models. Besides, ablation results are listed to verify the effectiveness of every module. Finally, we analyse our model about hyper-parameters and quality comprehensively.

\subsection{Dataset}
\textbf{ActivityCaption~\cite{regneri2013grounding}.} The dataset contains over 19k videos with diverse contents and their average duration is about 2 minutes. Following the standard split in~\cite{zhang2019cross,zhang2019learning}, there are about 37k, 18k and 17k sentence-moment pairs used for training, validation and testing respectively. ActivityCaption is the largest video moment retrieval benchmark with diverse contexts. Since the test set is not public, we regard a partial validation set as the test set in our experiments.

\textbf{Charades-STA~\cite{gao2017tall}.} The dataset is built on the Charades dataset~\cite{sigurdssonhollywood} which contains video-level paragraph descriptions and temporal action annotations.
Then Gao et al.\cite{gao2017tall} apply a semi-automatic approach to decompose video-level descriptions into short sentences for moments. Charades-STA contains 10k videos of indoor activities and their duration is about 30 seconds on average. There are around 12k moment-sentence pairs for training and 4k pairs for testing.

\textbf{DiDeMo~\cite{hendricks2017localizing}.} The dataset consists of 10k videos and the duration of each video is 25-30 seconds. It approximately contains 33k moment-sentence pairs for training, 4k for validation and 4k for testing. Specially, each video is divided into six five-second segments in DiDeMo and the target moment contains one or more consecutive segments. Therefore, there are only 21 moment candidates while ActivityCaption and Charades-STA datasets allow arbitrary temporal boundaries.

\begin{table}[t]
	\centering
	\caption{Performance Evaluation Results on ActivityCaption 
		($n \in \{1,5\}$ and  $m \in \{0.1,0.3,0.5\}$). }
	\label{table:activity}
	\renewcommand{\arraystretch}{1.3}
	\renewcommand\tabcolsep{2.5pt} 
	\scalebox{1}{
		\begin{tabular}{c|ccc|ccc}
			\hline
			\multirow{2}{*}{Method} & \multicolumn{3}{c|}{R@1} & \multicolumn{3}{c}{R@5} \\
			&  IoU=0.1&IoU=0.3 & IoU=0.5 & IoU=0.1& IoU=0.3 &  IoU=0.5 \\
			\hline
			\hline
			\multicolumn{7}{c}{fully-supervised methods}  \\
			\hline
			TGN~\cite{chen2018temporally}&-&43.81&27.93
			&-&54.56&44.20\\  
			QSPN~\cite{xu2019multilevel} &-&45.30&27.70
			&-&75.70&59.20\\
			2D-TAN~\cite{zhang2019learning}&-&59.45&44.51
			&-&85.53&77.13\\  
			\hline
			\hline
			\multicolumn{7}{c}{weakly-supervised methods}  \\
			\hline
			WS-DEC~\cite{duan2018weakly}&62.71&41.98&23.34
			&-&-&-\\
			WSLLN~\cite{gao2019wslln} &{\bf 75.40}&42.80&22.70
			&-&-&-\\
			CTF~\cite{chen2020look}&{74.20}&44.30&23.60
			&-&-&-\\
			SCN~\cite{lin2019weakly}&{71.48}&{47.23}&{29.22}
			&{90.88}&{71.45}&{55.69}\\
			RTPEN~(our)&{66.57}&{\bf 49.91}&{\bf 29.82}
			&{\bf 96.83}&{\bf 87.83}&{\bf 75.21}\\
			\hline
		\end{tabular}
	}
\end{table}

\begin{table}[t]\normalfont
	\centering
	\caption{Performance Evaluation Results on Charades-STA 
		($n \in \{1,5\}$ and  $m \in \{0.3,0.5,0.7\}$). }
	\label{table:charades}
	\renewcommand{\arraystretch}{1.3}
	\renewcommand\tabcolsep{2.5pt} 
	\scalebox{1}{
		\begin{tabular}{c|ccc|ccc}
			\hline
			\multirow{2}{*}{Method} & \multicolumn{3}{c|}{R@1} & \multicolumn{3}{c}{R@5} \\
			&  IoU=0.3&   IoU=0.5 &  IoU=0.7 & IoU=0.3&   IoU=0.5 &  IoU=0.7\\
			\hline
			\hline
			\multicolumn{7}{c}{fully-supervised methods}  \\
			\hline
			VSA-RNN~\cite{gao2017tall}&-&10.50&4.32
			&-&48.43&20.21\\
			VSA-STV~\cite{gao2017tall}&-&16.91&5.81
			&-&53.89&23.58\\
			CTRL~\cite{gao2017tall}&-&23.63&8.89
			&-&58.92&29.52\\
			QSPN~\cite{xu2019multilevel} &54.70&35.60&15.80
			&95.60&79.40&45.40\\
			2D-TAN~\cite{zhang2019learning}&-&39.81&23.25
			&-&79.33&52.15\\            
			\hline
			\hline
			\multicolumn{7}{c}{weakly-supervised methods}  \\
			\hline
			TGA~\cite{mithun2019weakly}&32.14&19.94&8.84
			&86.58&65.52&33.51\\
			CTF~\cite{chen2020look}&39.80&27.30&12.90
			&-&-&-\\
			SCN~\cite{lin2019weakly}&{42.96}&{23.58}&{9.97}
			&{95.56}&{71.80}&{38.87}\\
			RTPEN~(our) &{\bf 59.60}&{\bf 32.91}&{\bf 14.02}
			&{\bf 97.56}&{\bf 79.92}&{\bf 44.24}\\
			\hline
		\end{tabular}
	}
\end{table}

\subsection{Evaluation Criteria}
Following the widely-used setting in literature~\cite{gao2017tall,hendricks2017localizing}, we adopt {\bf R@n, IoU=m} as the criteria for ActivityCaption and Charades-STA, and we evaluate models by {\bf Rank@1}, {\bf Rank@5} and {\bf mIoU} for DiDeMo. 
Concretely, we calculate the Intersection over Union(IoU) between our predicted moments and ground truth, {\bf R@n, IoU=m} is the percentage of at least one of the top-n retrievals having the IoU $>$ m. And the {\bf mIoU} is the average IoU of the top-1 retrieval over all testing samples. 
For DiDeMo, since there are only 21 moment candidates in each video, {\bf Rank@1} or {\bf Rank@5} ,means the percentage of testing samples which correct moment is ranked as top-1 or among top-5.

\subsection{Implementation Details}
In this section, we introduce some implementation details of our RTPEN model.

\textbf{Data Preprocessing.} 
For raw videos and sentences, we first transform them into feature vectors. To have a fair comparison, we use the same visual features as previous methods~\cite{gao2017tall,hendricks2017localizing,zhang2019cross}. Concretely, we extract C3D features for ActivityCaption and Charades-STA datasets and apply VGG16 and optical flow features for DiDeMo. To reduce the complexity of our model, we shorten the feature sequence using temporal average pooling with the stride 8 and 4 for ActivityCaption and Charades-STA, respectively. As in~\cite{hendricks2017localizing}, we compute the average feature for each fixed five-second segment for DiDeMo.
As for language queries, we split the sentence into word and punctuation list, to reduce the impact of query noise and improve efficiency, we next discard the words not in  Glove~\cite{pennington2014glove} vocabulary and take the first $MaxSeq$ elements, where the $MaxSeq$ is set to 25, 20 and 20 for ActivityCaption, Charades-STA and DiDeMo respectively. Then we apply the pre-trained 300-d Glove embedding to extract semantic features for each word token.

\begin{table}[t]\normalfont
	\centering
	\caption{Performance Evaluation Results on DiDeMo. }
	\label{table:didemo}
	\renewcommand{\arraystretch}{1.3}
	\renewcommand\tabcolsep{4.0pt} 
	\scalebox{1}{
		\begin{tabular}{c|c|ccc}
			\hline
			{Method}&  Input&   Rank@1 &  Rank@5 &  mIoU\\
			\hline
			\hline
			\multicolumn{5}{c}{fully-supervised methods}  \\
			\hline
			MCN~\cite{hendricks2017localizing}&RGB&13.10&44.82&25.13\\
			TGN~\cite{chen2018temporally}&RGB&24.28&71.43&38.62\\
			\hline
			MCN~\cite{hendricks2017localizing}&Flow&18.35&56.25&31.46\\
			TGN~\cite{chen2018temporally}&Flow&27.52&76.94&42.84\\
			\hline
			MCN~\cite{hendricks2017localizing}&RGB+Flow&28.10&78.21&41.08\\
			TGN~\cite{chen2018temporally}&RGB+Flow&28.23&79.26&42.97\\
			\hline
			\hline
			\multicolumn{5}{c}{weakly-supervised methods}  \\
			\hline
			WSLLN~\cite{gao2019wslln} &RGB&19.40&53.10&25.40\\
			RTPEN~(our) &RGB&{\bf 20.18}&{\bf 60.37}&{\bf 28.21}\\
			\hline
			WSLLN~\cite{gao2019wslln} &Flow&18.40&54.40&27.40\\
			RTPEN~(our) &Flow&{\bf 20.89}&{\bf 60.16}&{\bf 30.98}\\
			\hline
			TGA~\cite{mithun2019weakly}&RGB+Flow&12.19&39.74&24.92\\
			RTPEN~(our) &RGB+Flow&{\bf 21.54}&{\bf 62.94}&{\bf 30.97}\\
			\hline
		\end{tabular}
	}
\end{table}

\begin{table*}[t]
	\centering
	\caption{Ablation results about the two-branch architecture, filter details and center-based proposal method.}
	\label{table:design}
	\renewcommand{\arraystretch}{1.3}
	\renewcommand\tabcolsep{3.5pt} 
	\scalebox{1}{
		\begin{tabular}{c|ccc|ccc|ccc|ccc}
			\hline
			\multirow{3}{*}{Method} & \multicolumn{6}{c|}{ActivityCaption} & \multicolumn{6}{c}{Charades-STA} \\
			\cline{2-13}
			& \multicolumn{3}{c|}{R@1} & \multicolumn{3}{c|}{R@5} & 
			\multicolumn{3}{c|}{R@1} & \multicolumn{3}{c}{R@5} \\
			& IoU=0.1&   IoU=0.3 &  IoU=0.5 & IoU=0.1&   IoU=0.3 &  IoU=0.5  &			
			IoU=0.3&   IoU=0.5 &  IoU=0.7 & IoU=0.3&   IoU=0.5 &  IoU=0.7\\
			\hline
			\hline
			\multicolumn{13}{c}{The Two-Branch Architecture} \\
			\hline
			w/o. filter
			&72.14& 45.05& 27.89
			&93.07&79.23&62.23
			&56.12&30.24&11.70
			&96.01&71.44&40.13\\
			w/o. erasing 
			&73.73&49.77&29.63
			&93.89&79.89&60.56
			&60.04&32.36&13.24
			&97.48&71.85&41.18\\
			w/o. parameter sharing   
			&{\bf 81.75}&47.25&23.34
			&92.35&75.97&58.22
			&{33.25}&12.94&5.37
			&81.93&46.28&19.26\\
			full model &{66.57}&{\bf 49.91}&{\bf 29.82}
			&{\bf 96.83}&{\bf 87.83}&{\bf 75.21}
			&{\bf 59.60}&{\bf 32.91}&{\bf 14.02}
			&{\bf 97.56}&{\bf 79.92}&{\bf 44.24}\\
			\hline
			\hline
			\multicolumn{13}{c}{The Filter Design} \\
			\hline
			visual-only scoring  
			&70.14&45.92&28.04
			&94.52&80.83&65.51
			&{57.23}&{30.28}&{12.83}
			&{96.81}&{71.87}&{40.82}\\
			w/o. NetVALD
			&{\bf 71.91}&46.39&28.81
			&95.11&82.95&67.29
			&{58.98}&32.04&13.21
			&97.08&72.23&41.28\\
			full model 
			&{66.57}&{\bf 49.91}&{\bf 29.82}
			&{\bf 96.83}&{\bf 87.83}&{\bf 75.21}
			&{\bf 59.60}&{\bf 32.91}&{\bf 14.02}
			&{\bf 97.56}&{\bf 79.92}&{\bf 44.24}\\
			\hline
			\hline
			\multicolumn{13}{c}{The Proposal Method} \\
			\hline
			all-proposal
			&{\bf 82.12}&48.52&21.53
			&93.72&79.89&66.82
			&{57.92}&{30.75}&{12.16}
			&{96.63}&{71.34}&{40.42}\\
			top-k proposal	
			&71.56&47.38&28.64
			&95.23&81.39&68.23
			&{58.19}&{32.01}&{12.82}
			&{96.81}&{74.21}&{41.63}\\
			full model (center-based) 
			&{66.57}&{\bf 49.91}&{\bf 29.82}
			&{\bf 96.83}&{\bf 87.83}&{\bf 75.21}
			&{\bf 59.60}&{\bf 32.91}&{\bf 14.02}
			&{\bf 97.56}&{\bf 79.92}&{\bf 44.24}\\
			\hline
			
		\end{tabular}
	}
\end{table*}

\textbf{Model Setting.} In RTPEN, the dimension of almost parameter matrices and bias is set to 512, including the ${\bf W}^{c}$, ${\bf b}^{c}$ in the NetVLAD, ${\bf W}^{m}_1$, ${\bf W}^{m}_2$ and ${\bf b}^{m}$ in the cross-modal interaction unit and so on. We set the dimension of the hidden state of each direction in Bi-GRU to 256, and the dimension of trainable center vectors is 512. 
During 2D temporal map construction, we fill all locations $[a, b]$ if $a <= b$ for DiDeMo. But for Charades-STA, we add another limitation $(b-a) \mod 2 = 1$. And for ActivityCaption, we only fill the location $[a, b]$ if $(b-a) \mod 8 = 0$. The
sparse sampling avoids much computational cost. And we set the convolution kernel size $K$ to 5, 3 and 1 for ActivityCaption, Charades-STA and DiDeMo respectively.
In the center-based proposal approach, the positive/negative proposal number $T$ is set to 16 for ActivityCaption, 32 for Charades-STA and 6 for DiDeMo.
During training, we set $\lambda_1$--$\lambda_5$ to 0.1, 1, 0.1, 0.01 and 0.01 respectively. For margin-based triplet loss, the margin $\Delta_{intra}$ and $\Delta_{inter}$ are set to 0.4 and 0.6 respectively. We use an Adam optimizer~\cite{duchi2011adaptive} with the initial learning rate $10^{-4}$ and weight decay $10^{-7}$. During inference, we apply the non-maximum suppression (NMS) with the threshold 0.55 while we need to select multiple moments.

\begin{figure}[!t]
	\setlength\tabcolsep{0.1 pt}
	\renewcommand{\arraystretch}{1}
	\begin{tabular}{ccc}
		\centering
		\includegraphics[height=0.19\textwidth,width=0.242\textwidth]{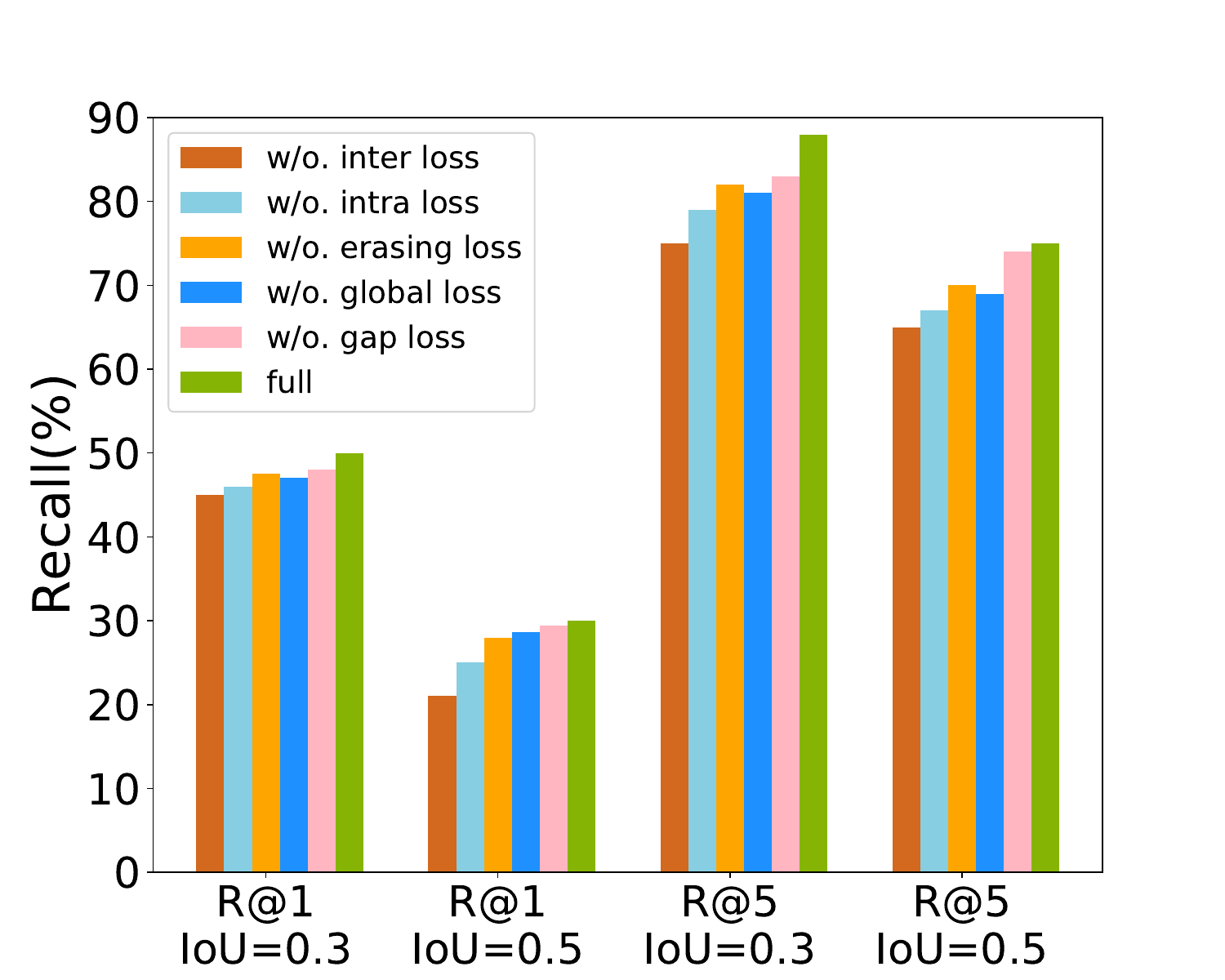} & \ \ & \includegraphics[height=0.19\textwidth,width=0.242\textwidth]{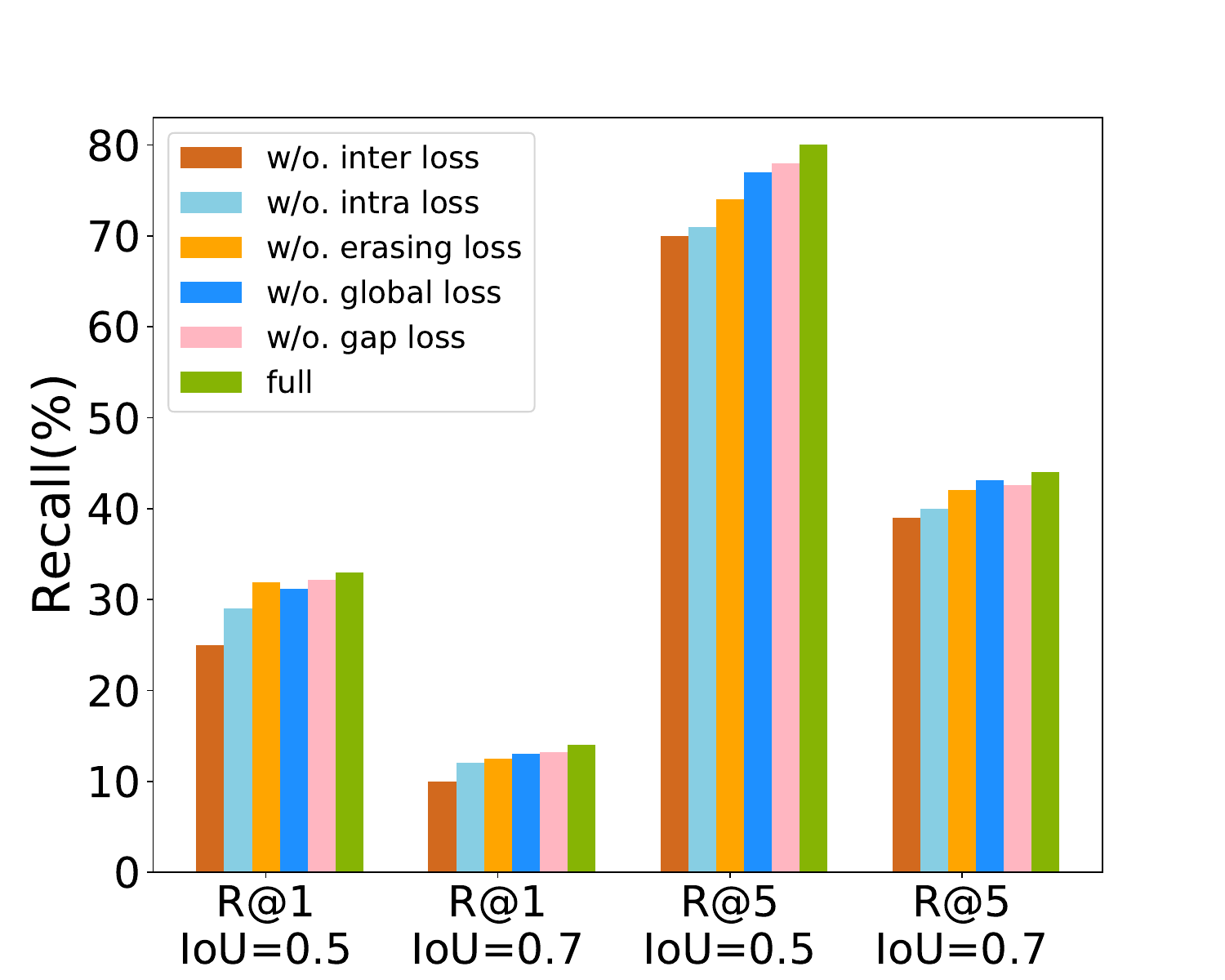}\\
		\footnotesize {(a) ActivityCaption }& &\footnotesize {(b) Charades-STA}
	\end{tabular}
	\caption{Ablation Results of the Multi-Task Losses.}
	\label{fig:loss} 
\end{figure}

\subsection{Comparison to State-of-the-Art Methods} 
We compare our RTPEN method with existing state-of-the-art methods, including the fully-supervised and weakly-supervised approaches.

\textbf{Supervised Method.} Early approaches VSA-RNN~\cite{gao2017tall}, VSA-STV~\cite{gao2017tall}, CTRL~\cite{gao2017tall} and MCN~\cite{hendricks2017localizing} project the features of candidate moments and language features into a common semantic space for correlation estimation. From a holistic perspective, TGN~\cite{chen2018temporally} builds the frame-by-word interaction by RNN. And QSPN~\cite{xu2019multilevel} integrates visual and textual features early and re-generate language descriptions as an auxiliary task.
Further, 2D-TAN~\cite{zhang2019learning} captures the temporal relations between adjacent video moments by a 2D temporal map.
\textbf{Weakly-Supervised Method.} WS-DEC~\cite{duan2018weakly} regards the weakly-supervised moment retrieval and video event captioning as dual problems, and develop a cycle system. Under the MIL-based framework, TGA~\cite{mithun2019weakly} uses the text-guided video attention distribution to detect the target moment, WSLLN~\cite{gao2019wslln} apply a alignment module and a detection module simultaneously to boost the performance, and CTF~\cite{chen2020look} retrieve the moment in a coarse-to-fine two-stage manner.
Different from MIL-based methods, SCN~\cite{lin2019weakly} ranks the moment proposals by a sentence reconstruction reward.

\begin{figure}[!t]
	\setlength\tabcolsep{0.1 pt}
	\renewcommand{\arraystretch}{1}
	\begin{tabular}{ccc}
		\centering
		\includegraphics[height=0.19\textwidth,width=0.242\textwidth]{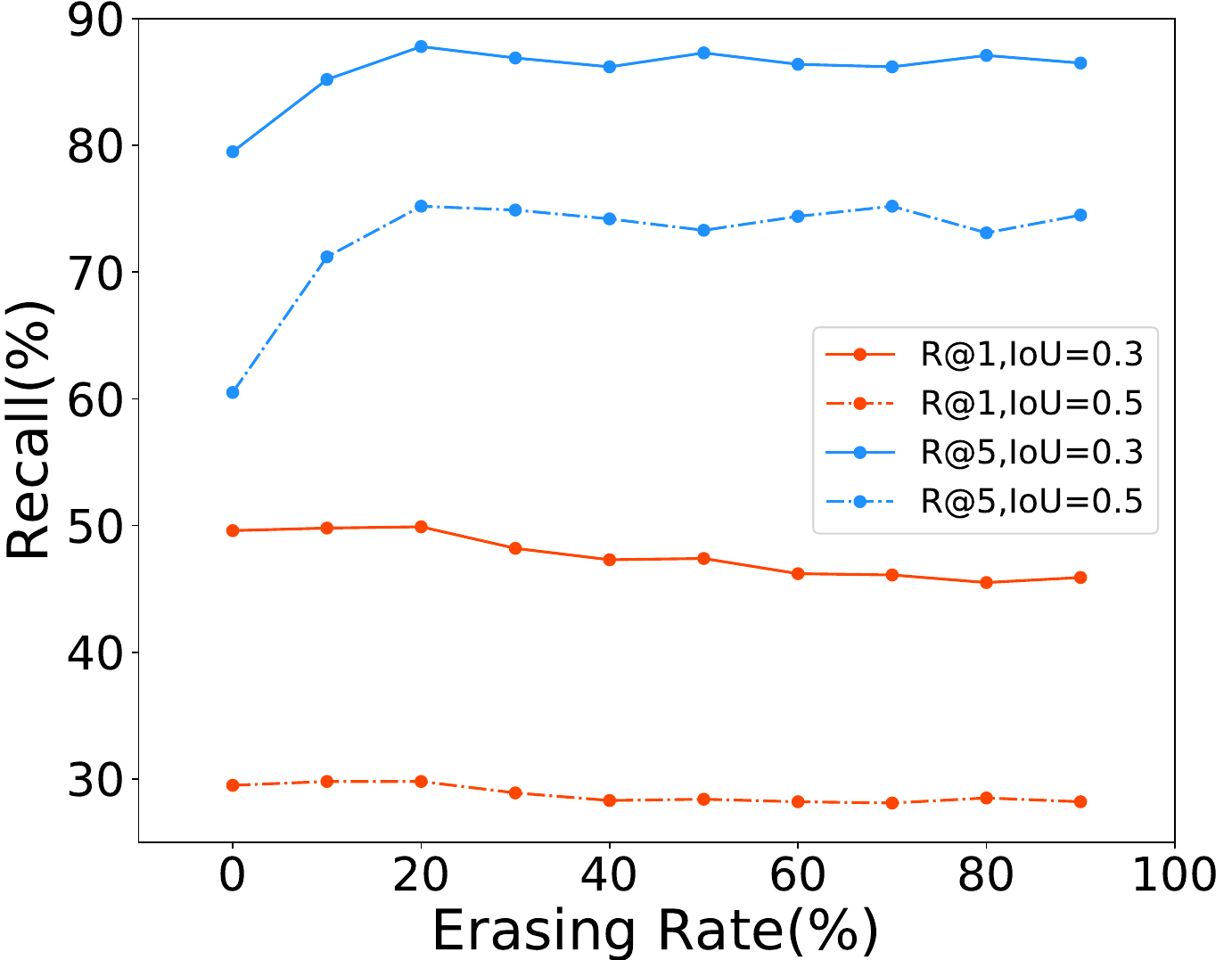} & \ \ & \includegraphics[height=0.19\textwidth,width=0.242\textwidth]{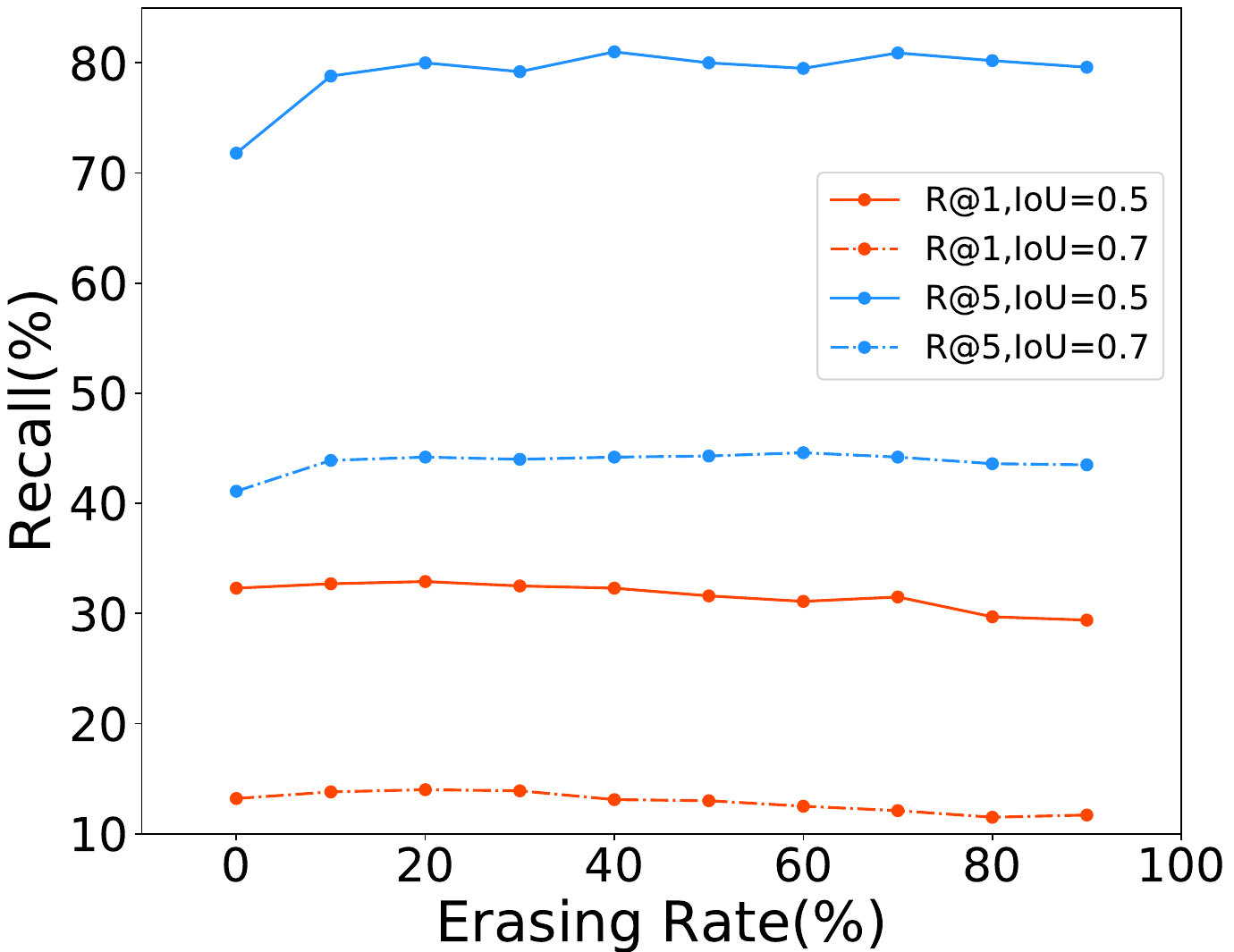}\\
		\footnotesize {(b) ActivityCaption}& &\footnotesize {(a) Charades-STA }
	\end{tabular}
	\caption{Effect of the Erasing Rate 
		on ActivityCaption and Charades-STA.}
	\label{fig:erasing rate} 
\end{figure}

The overall experience results on three large-scale datasets are displayed in  Table~\ref{table:activity}, Table~\ref{table:charades} and Table~\ref{table:didemo}, where we set $n \in \{1,5\}, m \in \{0.1,0.3,0.5\}$ for ActivityCaption and $n \in \{1,5\}, m \in \{0.3,0.5,0.7\}$ for Charades-STA. The results reveal some interesting points:
\begin{itemize}
	\item On three datasets, our RTPEN method almost achieves the best weakly-supervised performance, especially on Charades-STA and DiDeMo, verifying the effectiveness of our two-branch framework with the erasing mechanism and regularization strategy. 
	\item The R@5 of our RTPEN is extremely high, even exceeds all fully-supervised methods we compare to when IoU=0.3 and IoU=0.5 on ActivityCaption and Charades-STA respectively, which shows our dynamic erasing mechanism contributes to capturing the comprehensive video-sentence correspondence and retrieving more accurate moments. 
	
	\item The reconstruction-based method SCN outperforms MIL-based methods TGA, CTF and WSLLN on ActivityCaption and Charades-STA, but our RTPEN achieves a better performance than SCN, demonstrating our RTPEN with the intra-sample confrontment can effectively discover the plausible negative samples and improve the accuracy.

	\item Our RTPEN outperforms the state-of-the-art baselines using RGB, Flow and two-stream features on the DiDeMo dataset, which suggests our method is robust for diverse features.
\end{itemize}


%

\subsection{Ablation Study}
To better understand our model, we evaluate the influence of different losses in multi-task learning and explore the contribution of concrete designs of our RTPEN. 

\subsubsection{Ablation Study for the Multi-Task Loss}
We discard one loss from the multi-task loss at a time to build an ablation model, including \textbf{w/o. intra loss}, \textbf{w/o. inter loss} and so on. The results on ActivityCaption and Charades-STA are displayed in Fig.~\ref{fig:loss}, showing that the full model outperforms every ablation model on two datasets, 
which demonstrates all losses are indispensable --- the inter-sample and intra-sample losses effectively offer the supervision signals, meanwhile, the erasing loss can boost the cross-modal interaction and the regularized global and gap losses can promote the model performance. 

Further, we find the model~(w/o. inter loss) achieves the worst accuracy, fitting our perception that inter loss is the core of MIL-based methods, thus it is crucial. 
The model~(w/o. intra loss) and the model~(w/o. global loss) suffer the second largest declination, suggesting intra-sample confrontment is significant for weakly-supervised moment retrieval and filtering out irrelevant moments is important to model training. 

\begin{figure}[!t]
	\setlength\tabcolsep{0.1 pt}
	\renewcommand{\arraystretch}{1}
	\begin{tabular}{ccc}
		\centering
		\includegraphics[height=0.19\textwidth,width=0.242\textwidth]{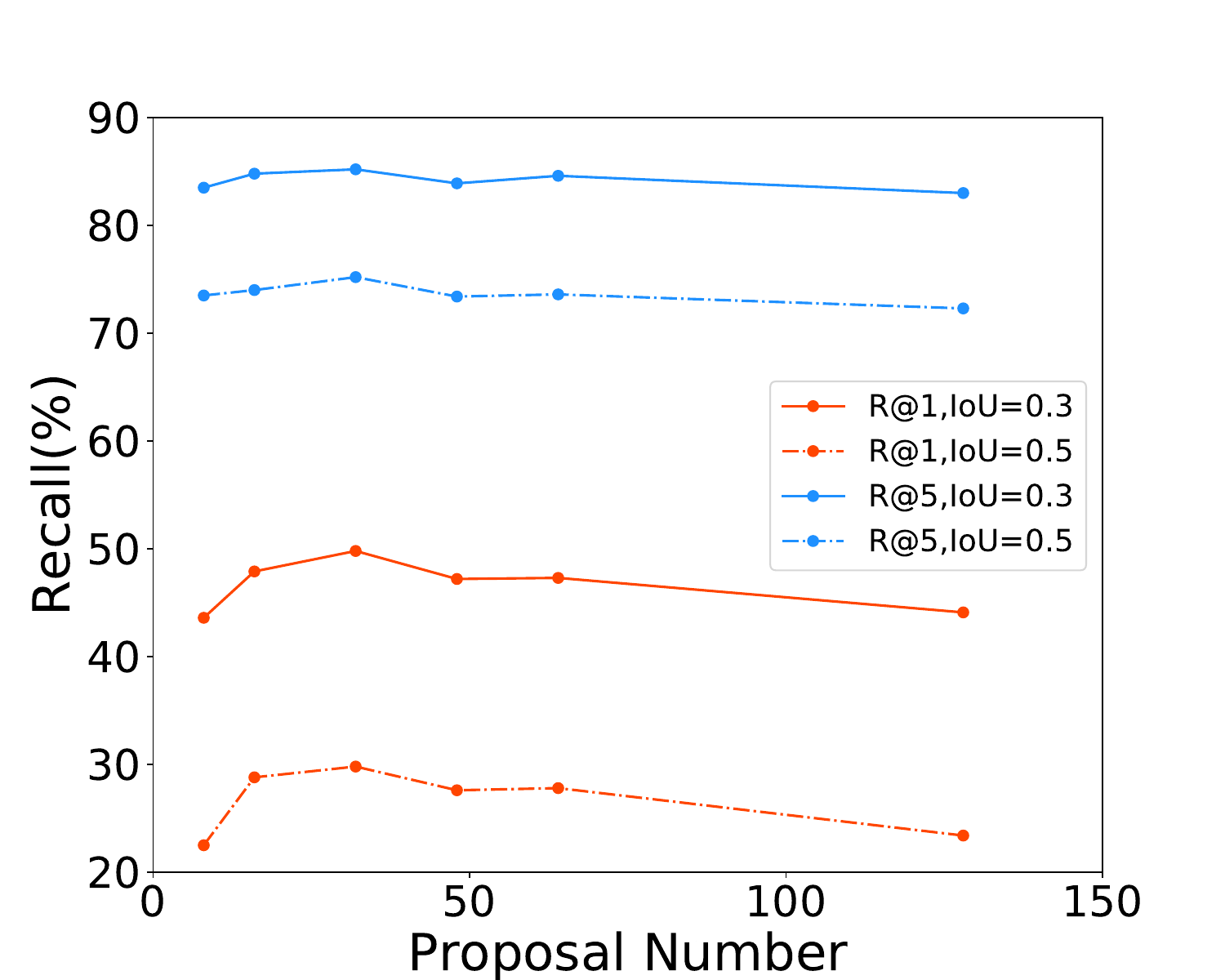} & \ \ & \includegraphics[height=0.19\textwidth,width=0.242\textwidth]{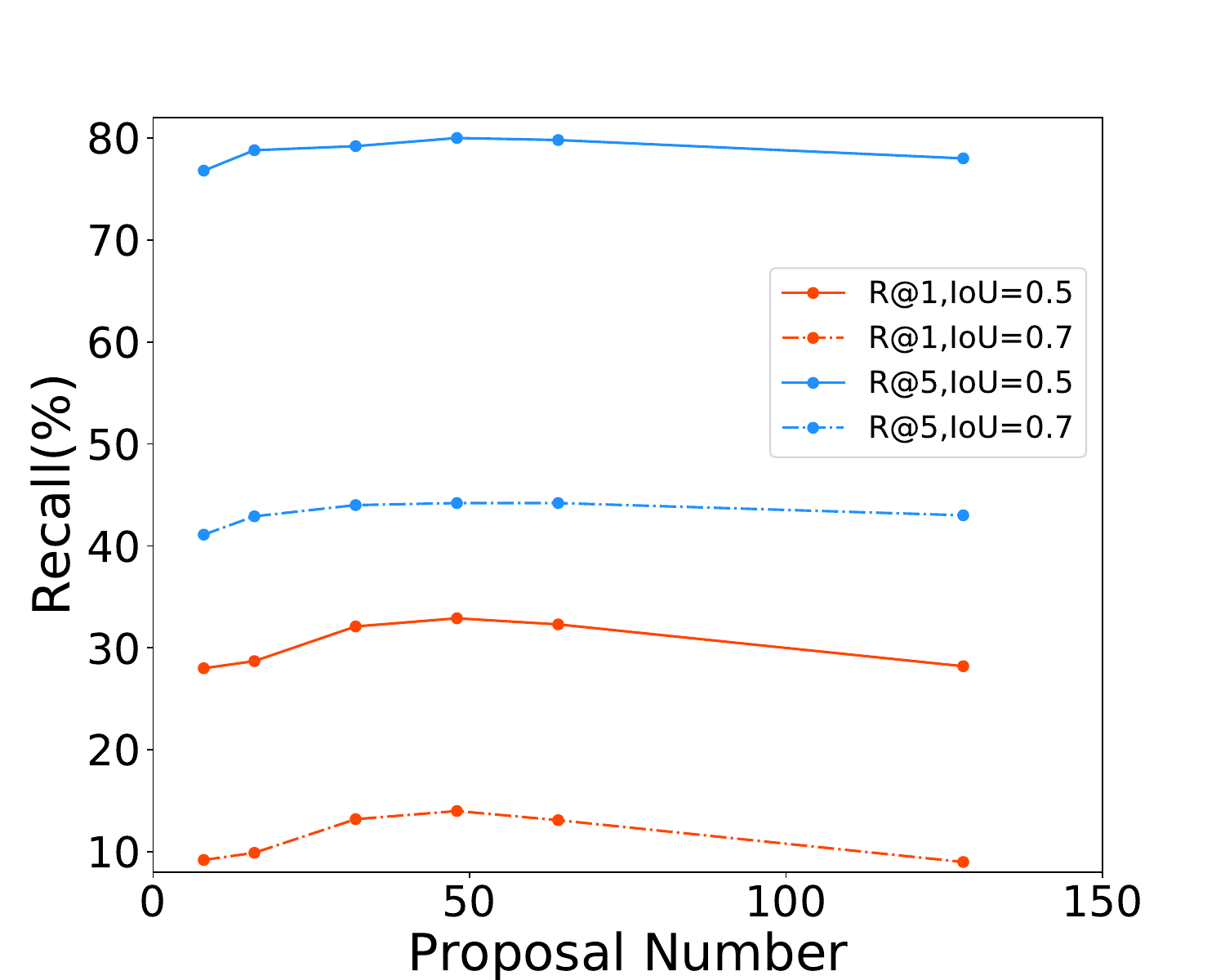}\\
		\footnotesize {(a) ActivityCaption}& &\footnotesize {(b) Charades-STA}
	\end{tabular}
	\caption{Effect of the Proposal Number on ActivityCaption and Charades-STA.}
	\label{fig:hyper} 
\end{figure}

\subsubsection{Ablation Study for the Model Design}
We next verify the effectiveness of our model design, including the two-branch architecture, filter designs and center-based proposal method. We note the cross-modal interaction unit~\cite{zhang2019cross} and 2D temporal map~\cite{zhang2019learning} are mature techniques which do not need further ablation.
\begin{itemize}
	
	\item \textbf{Two-Branch Architecture.} 
	We remove the crucial visual filter and retain a single branch with erasing mechanism to perform the conventional MIL-based training without the intra-sample loss as \textbf{w/o. filter}. Then we keep the entire framework, withdrawing the erasing module in the enhanced branch as \textbf{w/o. erasing}, discarding the parameter sharing between two branches as \textbf{w/o. parameter sharing}.
	\item \textbf{Filter Design.} We discard the cross-modal estimation and produce the visual score distribution by only frame features as \textbf{w/o. visual-only scoring}. And we remove the NetVALD and directly apply the language features during cross-modal estimation as \textbf{w/o. NetVALD}.
	\item \textbf{Proposal Method.} We discard the center-based proposal method and sample all candidate moments during training as \textbf{all-proposal}. And we utilize a top-k proposal method instead of the center-based proposal strategy as \textbf{top-k proposal}, where we select $T$ moments with high proposal scores directly.
\end{itemize}

\begin{figure}[t]
	\centering
	\includegraphics[width=1.0\columnwidth]{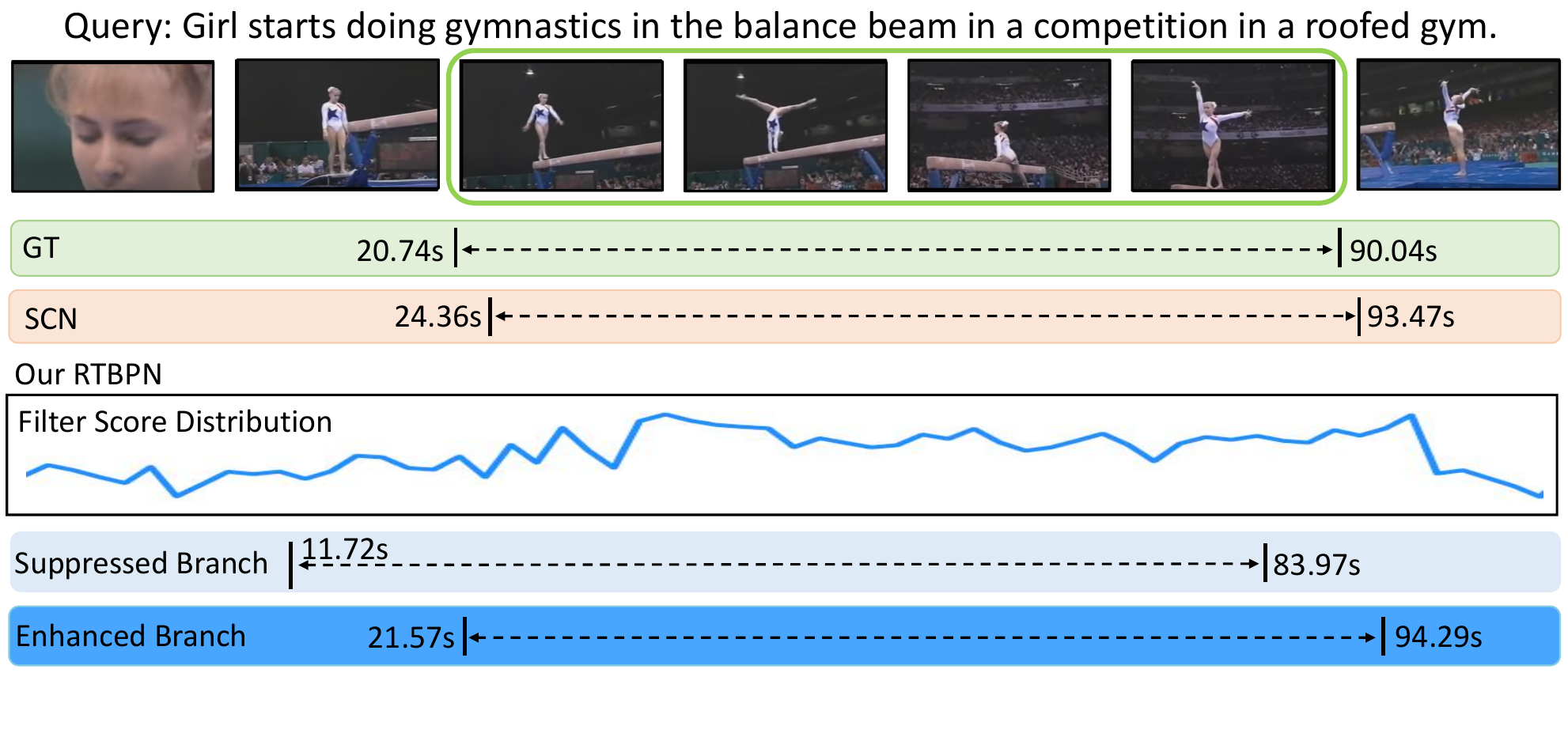} 
	\includegraphics[width=1.0\columnwidth]{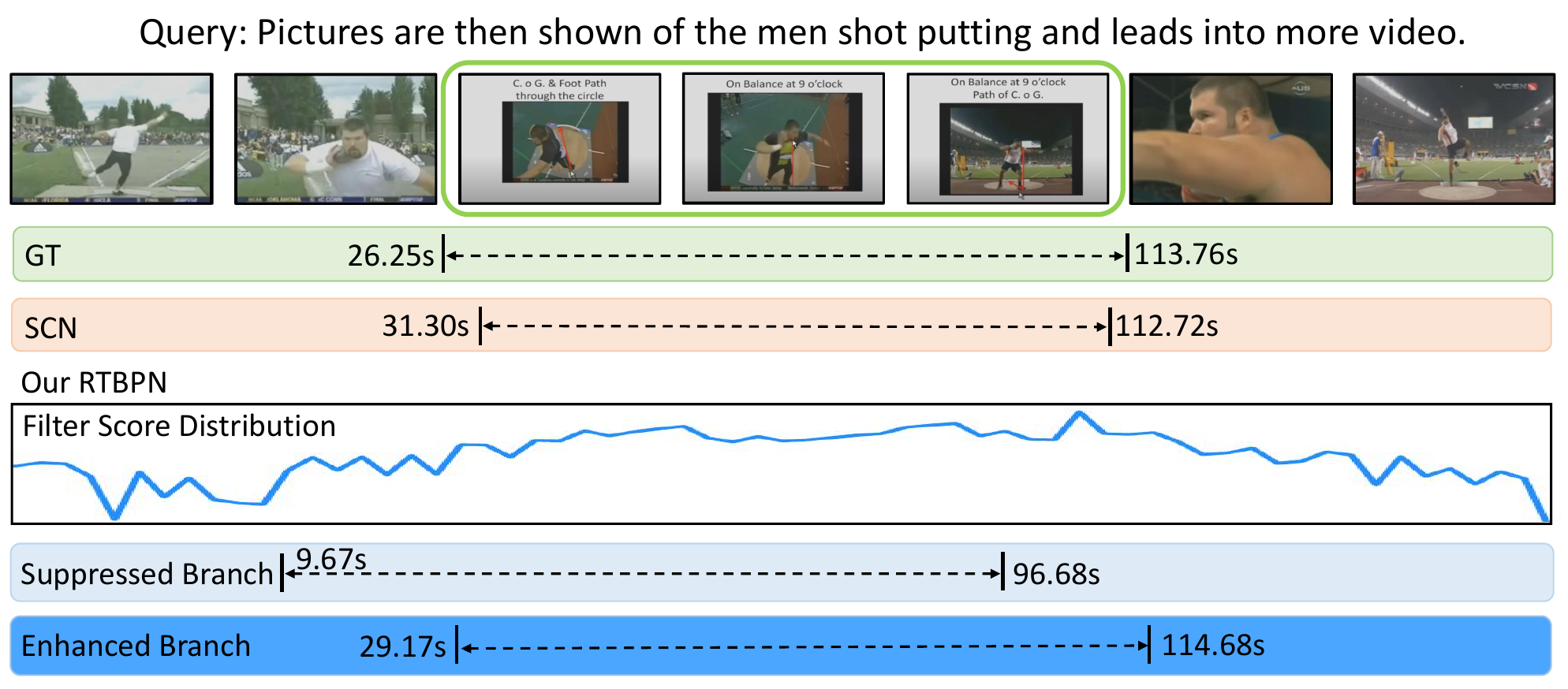} 
	\caption{Qualitative Examples on ActivityCaption and Charades-STA.}
	\label{fig:example_act}
\end{figure}

The ablation results on ActivityCaption and Charades-STA datasets are reported in Table~\ref{table:design} and we can  achieve the following conclusions:
\begin{itemize}
	\item The model~(w/o. filter) and model~(w/o. parameter sharing) 
	have performance degradation severely than the full model, demonstrating that the two-branch architecture with the language-aware visual filter can develop the intra-sample confrontment and improve the model performance. The performance of model~(w/o. erasing) also decreases a lot, especially on R@5, which shows the dynamic erasing mechanism contributes to capturing the comprehensive query semantics for more accurate retrieval, and the parameter sharing is vital to make two branches produce critical proposals for sufficient confrontments.
	\item The full model achieves better results than model~(visual-only scoring) and model~(w/o. NetVALD), which suggests that the cross-modal estimation with language semantic information can generate a more reasonable score distribution than visual-only scoring. And the NetVALD can further promote the cross-modal estimation by introducing an intermediately semantic space for videos and texts.
	\item As for the proposal method, the model with the center-based strategy can outperform the model~(all-proposal) and model~(top-k proposal). It verifies our center-based proposal method can discover a series of correlative moments effectively for MIL-based inter-sample and intra-sample training.
	\item Actually, some ablation models, e.g. model~(visual-only scoring), model~(w/o. erasing) and model~(top-k proposal), still yield better performance than state-of-the-art baselines, proving our RTPEN network is robust and does not rely on a key component. 
\end{itemize}

\begin{figure}[t]
	\centering
	\includegraphics[width=1.0\columnwidth]{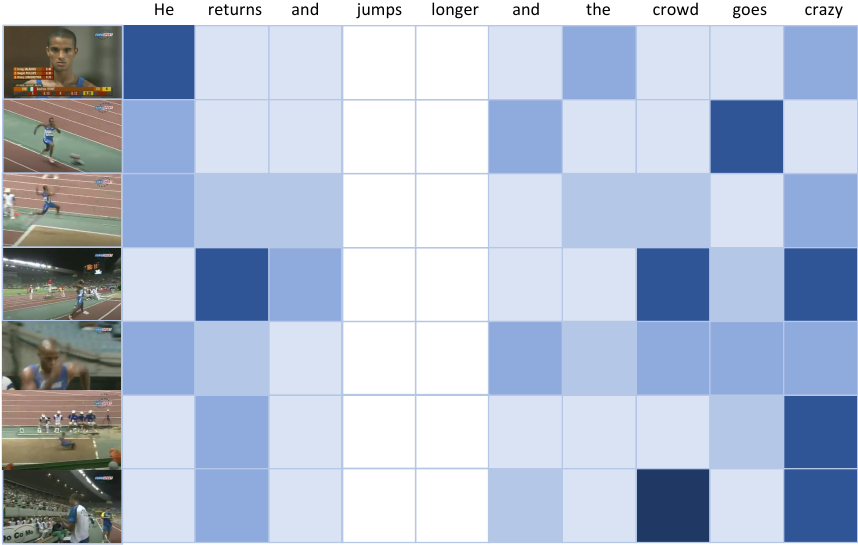} 
	\caption{Cross-Modal Attention Heat Map from Erasing Enhanced Branch.}
	\label{fig:attention last}
\end{figure}

\subsection{Hyper-Parameters Analysis}
In RTPEN model, the number of selected positive/negative proposals $T$ is a significant hyper-parameter. Thus, we further explore its effect by varying the proposal number. Specifically, we set $T$ to 8, 16, 32, 48, 64, 128 on ActivityCaption and Charades-STA and report our experiment results in Figure~\ref{fig:hyper}, where we select both R@1 and R@5 as the evaluation criteria when IoU=0.3 and IoU=0.5 for ActivityCaption and IoU=0.5 and IoU=0.7 for Charades-STA. From a holistic view, the curve of R@1 is more fluctuant than R@5, which means $T$ is influential for retrieving the most relevant moment.
Further, we find that and the model achieves the best performance on ActivityCaption and Charades-STA when $T$ is set to 32 and 48 respectively, which is approximately the 25\% of all proposals on each dataset. Because too many proposals may introduce irrelevant moments during training and decrease the performance. And too few proposals will fail to develop sufficient confrontments between crucial moments and lead to the poor performance. 
Moreover, the trends of the influence of the proposal number $T$ on two datasets are similar, demonstrating it is insensitive to different datasets.

Further, to probe into what percentage of erasing is appropriate in our dynamic erasing mechanism, we set erasing rate $E$ from 10\% to 90\% with the step 10\%, and $E=0$ means w/o. erasing mechanism, the results are shown in Figure \ref{fig:erasing rate}. We can find the models achieves better performance than larger rates intuitively when $E$ are 10\% and 20\%, since the primary model with sharp attention distribution (as Fig.~\ref{fig:attention} shown) tends to focus on only about 10\%-20\% words in our experience, erasing these words is enough to encourage the model to excavate the subdominant language semantics. 
As the erasing rate keeps increasing, our model maintains high R@5 performance, and it has the R@1 performance degradation to the model (w/o. erasing), which still performs well.

\begin{figure}[t]
	\centering
	\includegraphics[width=1.0\columnwidth]{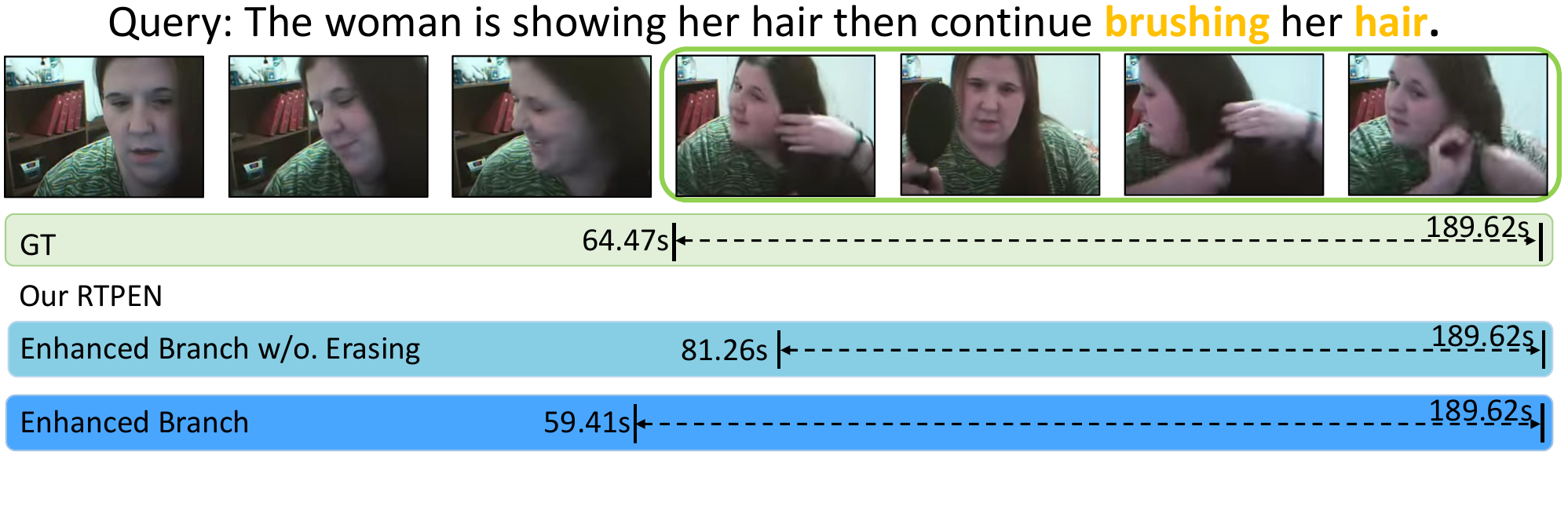} 
	\includegraphics[width=1.0\columnwidth]{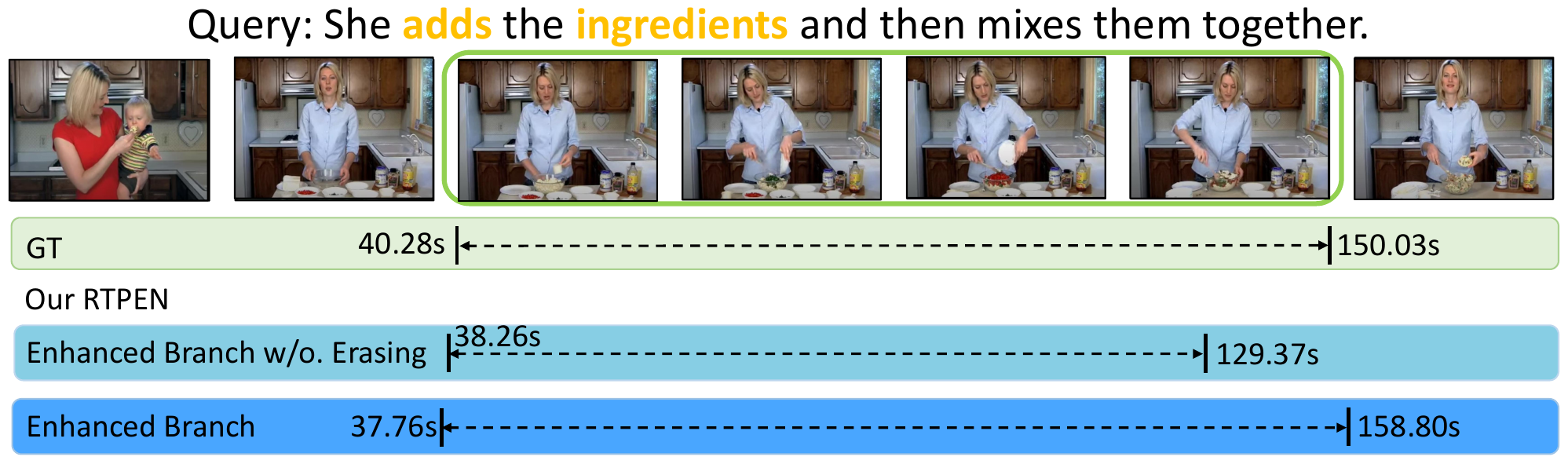} 
	\caption{Qualitative Examples about Erasing Mechanism on ActivityCaption and Charades-STA.}
	\label{fig:example_erase}
\end{figure}

\subsection{Qualitative Analysis}
To validate the effectiveness of our RTPEN qualitatively, we first display two typical examples on ActivityCaption and Charades-STA datasets in Fig.~\ref{fig:example_act}, where we show the retrieval result of the SCN baseline, the results from the enhanced branch and suppressed branch of RTPEN, and the score distribution from the language-aware visual filter.
By intuitive comparison, we find that our RTPEN can localize a more accurate moment than SCN from the enhanced branch, qualitatively verifying the effectiveness of our method. And we can observe that the language-relevant frames are given higher scores to than unnecessary ones by the filter. Based on the reasonable score distribution, the enhanced branch can localize the precise moment while the suppressed branch can only detect the relevant but not accurate moment as the plausible negative proposal. 

In addition, we visualize the attention heat map from erasing enhanced branch in Fig.~\ref{fig:attention last}, while Fig.~\ref{fig:attention} represents primary attention. After erasing the top 20\% focused words “jumps” and “longer”, the model reconcentrates on the other part of sentence query (“crowd” and “crazy”) which is neglected in primary branch before. Further, Fig.~\ref{fig:attention last} shows the retrieval results from the primary enhanced branch(w/o. erasing mechanism) and full enhanced branch on two datasets, where the yellow words in query are dominant in primary branch and are masked in erasing branch to capture complementary query semantics. We can find the moments localized from primary enhance branch are incomplete and have lower accuracy than full enhanced branch since the primary branch focuses on a few words exorbitantly and ignores complete sentence semantics to retrieve comprehensively. Therefore, by erasing mechanism, our model can capture complementary language semantics and comprehensive video-sentence relations to achieve better performance.

\section{Conclusion}
In this paper, we propose a novel Regularized Two-Branch Proposal Network with Eraseing Mechanism for weakly-supervised video moment retrieval. First we devise a language-aware visual filter to produce the enhanced and suppressed video streams, and then devise a sharable two-branch proposal module with erasing mechanism to generate positive and plausible negative moment proposals from the enhanced and suppressed streams respectively. 
Further, we apply the proposal regularization strategy to improve the model performance. The extensive experiments on three datasets demonstrate the effectiveness of our RTPEN model.



%





\ifCLASSOPTIONcaptionsoff
  \newpage
\fi



%


\newpage 

\bibliographystyle{IEEEtran.bst}
\bibliography{ref}

\end{document}